\documentclass[nofootinbib,twocolumn,preprintnumbers]{revtex4-1}
\pdfoutput=1
\usepackage{cancel}
\usepackage{amsmath,amsthm,amssymb,multirow,psfrag}
\usepackage{epsfig}
\usepackage{color}
\usepackage[compat=1.0.0]{tikz-feynman}
\usepackage[utf8]{inputenc}
\usepackage{slashed}
\usepackage[normalem]{ulem}
\usepackage{soul}
\usepackage{fixltx2e}
\usepackage{cancel}
\graphicspath{{./Figures/}}

\begin{document}

\def\lsim{\mathrel{\rlap{\lower4pt\hbox{\hskip1pt$\sim$}}
  \raise1pt\hbox{$<$}}}
\def\gsim{\mathrel{\rlap{\lower4pt\hbox{\hskip1pt$\sim$}}
  \raise1pt\hbox{$>$}}}
\newcommand{\vev}[1]{ \left\langle {#1} \right\rangle }
\newcommand{\bra}[1]{ \langle {#1} | }
\newcommand{\ket}[1]{ | {#1} \rangle }
\newcommand{\ev}{ {\rm eV} }
\newcommand{\kev}{{\rm keV}}
\newcommand{\mev}{{\rm MeV}}
\newcommand{\tev}{{\rm TeV}}
\newcommand{\mpl}{$M_{Pl}$}
\newcommand{\mw}{$M_{W}$}
\newcommand{\Ft}{F_{T}}
\newcommand{\Zparity}{\mathbb{Z}_2}
\newcommand{\BLambda}{\boldsymbol{\lambda}}
\newcommand{\met}{\;\not\!\!\!{E}_T}
\newcommand{\beq}{\begin{equation}}
\newcommand{\eeq}{\end{equation}}
\newcommand{\bea}{\begin{eqnarray}}
\newcommand{\eea}{\end{eqnarray}}
\newcommand{\nn}{\nonumber}
\newcommand{\gev}{{\mathrm GeV}}
\newcommand{\hc}{\mathrm{h.c.}}
\newcommand{\eps}{\epsilon}
\newcommand{\bwt}{\begin{widetext}}
\newcommand{\ewt}{\end{widetext}}
\newcommand{\draftnote}[1]{{\bf\color{blue} #1}}

\newcommand{\cO}{{\cal O}}
\newcommand{\cL}{{\cal L}}
\newcommand{\cM}{{\cal M}}

\newcommand{\fref}[1]{Fig.~\ref{fig:#1}} 
\newcommand{\eref}[1]{Eq.~\eqref{eq:#1}} 
\newcommand{\aref}[1]{Appendix~\ref{app:#1}}
\newcommand{\sref}[1]{Section~\ref{sec:#1}}
\newcommand{\tref}[1]{Table~\ref{tab:#1}}

\title{\LARGE{{\bf Triggering on Emerging Jets} }}
\author{{\bf {Dylan Linthorne and Daniel Stolarski}}}

\affiliation{
Ottawa-Carleton  Institute  for  Physics,  Carleton  University,\\
1125  Colonel  By  Drive,  Ottawa,  Ontario  K1S  5B6,  Canada
}

\email{
dylan.linthorne@carleton.ca \\
stolar@physics.carleton.ca \\}

\begin{abstract}
Confining dark sectors at the GeV scale can lead to novel collider signatures including those termed emerging jets with large numbers of displaced vertices. The triggers at the LHC experiments were not designed with this type of new physics in mind, and triggering can be challenging, especially if the mediator is relatively light and/or has quantum numbers such that additional jets are not automatically produced in each event. We show that the efficiency and the total event rate at current triggers can be significantly improved by considering initial state radiation of the events, with the largest increase in rate coming from simulation of two additional jets. We also explore possible new triggers that employ hit counts in different tracker layers as input into a machine learning algorithm. We show that these new triggers can have reasonably low background rates, and that they are sensitive to a wide range of new physics parameters even when trained on a single model.

\end{abstract}

\maketitle

\section{Introduction} 
\label{sec:intro} 

Confining hidden sectors with a confinement scale around the GeV scale have received significant attention for potential discovery at colliders~\cite{Schwaller:2015gea, Cohen_2015, Csaki:2015fba,Cohen_2017, Knapen_2017,Pierce:2017taw,Lichtenstein:2018kno,Cheng:2019yai} (for a recent review, see chapter 7 of~\cite{Alimena_2020_ds}), building on the seminal hidden valley work~\cite{Strassler:2006im,Strassler:2006ri,Han:2007ae}. 
Besides providing interesting signatures at colliders, they can also be motivated by asymmetric dark matter~\cite{Bai_2014,Lonsdale:2017mzg,Lonsdale:2018xwd} and by neutral naturalness~\cite{Craig:2015pha,Curtin:2015fna,Cheng:2016uqk,Kilic:2018sew}. This has led to several phenomenological studies at the LHC~\cite{Park:2017rfb,Cohen:2020afv,Mies:2020mzw,Knapen:2021eip}, flavour experiments~\cite{Renner:2018fhh}, and future proposed experiments~\cite{Alekhin:2015byh,Curtin:2018mvb}. There is also an experimental search~\cite{Sirunyan:2018njd} that puts direct limits on certain regions of parameter space.

If the dark confining sector has a mediator to the SM whose mass is much larger than the confining scale, then the lifetime of the lightest dark hadrons that are not stable will be parametrically larger than their inverse mass. One particularly interesting region of parameter space is where the lifetime of the decaying dark hadrons is in the mm--m range, leading to particularly spectacular signatures at the LHC's detectors~\cite{Schwaller:2015gea}. For example, if the dark gauge group is QCD-like, then when dark quarks are produced, they will shower and hadronize, producing dark jets. This is analogous to the production of ordinary quarks at a high energy lepton collider which then produce SM jets. Each of the dark hadrons will decay at a different position in the detector, and the energy of the dark jet will ``emerge'' into the detector, and this signature was thus termed an emerging jet~\cite{Schwaller:2015gea}. At distances long compared with the typical $c\tau$ of the dark hadrons, the dark jet will look like a QCD-jet. 

Motivated by the asymmetric dark matter scenario of~\cite{Bai_2014}, the work~\citep{Schwaller:2015gea} considered a scalar mediator $X$ that is charged under QCD and dark-QCD. That means that the dominant collider signal will be pair production of the mediator, and each collider event will contain two emerging jets and two SM jets. This is also the scenario experimentally constrained in~\cite{Sirunyan:2018njd}. Another well motivated possibility is the one considered in the original hidden valley literature~\cite{Strassler:2006im,Strassler:2006ri,Han:2007ae}: a vector mediator $Z'$ that couples to both quark and dark quark currents. A third possibility, the one expected in the neutral naturalness scenarios~\cite{Craig:2015pha,Curtin:2015fna,Cheng:2016uqk,Kilic:2018sew}, is the SM Higgs or another scalar as the mediator ($H$) to the dark sector. In both of the latter cases, we see two important differences between the original case of the $X$:
\begin{itemize}
\item The production cross section of dark hadrons is a free parameter and not set by pair production via SM QCD.
\item The production does not automatically include the associated production of SM jets.
\end{itemize}
These cases are therefore more challenging experimentally. 

A particularly important challenge with exotic signatures at the LHC is triggering. The event rate at the LHC is extremely high, and a trigger is employed to reduce the event rate by several orders of magnitude and attempt to record all the events of interest. The triggers were designed to be extremely efficient on many types of events and new physics models, but they are not designed for more exotic scenarios such as emerging jets. In the case of the $X$ mediator, if its mass is $\mathcal{O}$(TeV), then the associated jets in combination with the very high total energy in the event makes triggering relatively straightforward. 

In this work we study on the significantly more difficult case of triggering on the $Z'$ mediator, focusing on the relatively lower mass regime, $M_{Z'} \sim$ 100 GeV -- 1 TeV. We will take a two-pronged approach. First, we will consider how well current triggers can capture these events, and explore how the addition of initial state radiation can increase the efficiency. The so-called mono-$X$ strategy~\cite{Bai:2010hh,Goodman:2010ku} has been used extensively to search for invisible states, with a broad range of experimental searches including mono-jet~\cite{Aad:2021egl}, mono-photon~\cite{Sirunyan:2017ewk}, mono-$W$ and mono-$Z$~\cite{Sirunyan:2017hci}, placing constraints on many different types of models. In this work we show that both QCD and electroweak radiation can increase the trigger efficiency and increase the total number of events recorded.

The second strategy we employ is to consider new triggers using modern machine learning (ML) techniques. The landscape of machine learning applications within particle physics is becoming ever broader and more complex. Its utilities ranges from substructure classifications~\cite{Larkoski_2020}, such as jet discrimination~\cite{Metodiev_2017,Metodiev_2018}, to multivariate analysis techniques explored at both CMS and ATLAS as the LHC moves towards the energy and intensity frontiers.~\cite{doi:10.1146/annurev-nucl-101917-021019,Radovic}. 
We look to investigate novel triggers based on simple machine learning methods that can be applied to the triggering stream. Complementary studies include~\cite{Alimena_2020} where a deep neural-net (DNN) is implemented at L1 resulting in high trigger efficiencies for a HL-LHC 15 KHz target output, as well as the study of a more traditional trigger for displaced vertices~\cite{Gershtein:2020mwi}. CMS (ATLAS) has been slowly conducting analyses of this type on low level information reconstructed from the triggering stream under the guise of Data scouting (Online trigger analysis)~\citep{Mukherjee:2019anz, Anderson:2016ron, ATL-DAQ-PUB-2017-003}.

One significant downside to the use of ML techniques in particle physics is that the physical intuition of traditional analyses can be lost when using complicated non-linear cuts dictated by ML algorithms. It is often difficult to determine which aspects of events the algorithms are using to, for example, distinguish signal from background, which in turn makes it difficult to account for effects such as Monte Carlo mismodeling and detector uncertainties, although there is some recent progress~\cite{Faucett:2020vbu}. This is a significantly less important problem when considering triggers where the most important task is to get interesting events recorded quickly. One can then use a more careful offline analysis to gain physical insight and more completely characterize things like systematic errors.

In this work we use hits in the tracker as input into a potential new trigger discriminator. Tracking is traditionally not used in triggering because track reconstruction is too computationally time consuming~\citep{Owen:2302730,Khachatryan_2017}. We skirt this problem by not reconstructing tracks, but rather by simply counting hits in a region of the detector. This can be effective in distinguishing emerging jets because the dark sector particles will not leave hits but their decay products will. Therefore, if the dark sector particles lifetime $c\tau \sim$ mm -- m, an emerging jet will have increasing numbers of hits in detector layers further from the interaction point. QCD jets, on the other hand, will typically have the same number of hits in most layers because unstable SM hadrons will typically decay before hitting the first or second layer of the detector, with the exception of a few strange mesons. We will show that this type of observable fed into a machine learning algorithm can be an effective trigger for a wide class of model parameters. We explore the universality of such strategies and show that a trigger trained on one parameter point can be sensitive to a broad swath of parameter space.

The remainder of this paper is structured as follows: in Section~\ref{sec:model} we describe the concrete model we use for our analysis, and in Section~\ref{sec:event} we describe our simulation pipeline. In Section~\ref{sec:current} we describe how the mono-X strategy can be used to improve event collection with current triggers, and in Section~\ref{sec:ML} we outline how new triggers can also be used to explore new parameter space. Conclusions are given in Section~\ref{sec:conclusion}.

\section{Models for emerging jets}
\label{sec:model}

The study of generic hidden sectors at the LHC is an interesting and important question. For concreteness we will specify a class of models and leave the more general case to future work. We consider an $SU(N_d)$ gauge group with confinement scale $\Lambda_d \simeq$ GeV and $n_f$ flavours of vectorlike quarks with masses below confinement scale. The dark quarks are singlets under all SM gauge groups. If there is an accidental baryon number symmetry analogous to QCD, then the baryons of this sector could be dark matter~\cite{Bai_2014}. 

Unlike previous work which studied a scalar mediator~\cite{Schwaller:2015gea}, we consider a vector mediator, $Z'_\mu$ as in some of the original hidden valley literature~\citep{Strassler:2006im}, whose mass $M$ is larger than the dark confinement scale, $M \gg \Lambda_d$. The UV theory and mechanism to give mass to the $Z'$ is left unspecified, but we assume the additional states needed to not affect the phenomenology. This mediator couples to the quark and dark quark currents:
\begin{equation}\label{eqn:lagrangian}
\mathcal{L} \supset \frac{1}{2}M^2 Z'^{ \mu }Z'_{\mu} + Z'^{ \mu}  \big( g_q \, \overline{q}\gamma_\mu q + g_{d} \, \overline{Q} \gamma_\mu Q  \big) \, ,
\end{equation}
where $q$ are SM quarks, $Q$ are dark quarks, and $g_{q/{d}}$ are coupling constants. The $Z'$ is a singlet under SM and unbroken dark gauge groups, so gauge indices among the quarks and dark quarks are contracted in the standard way. For simplicity we have assumed flavour universality for both quarks and dark quarks with a universal coupling for each, and flavour indices are summed and not written. 

This model contains a rich spectrum of dark hadrons, with a multiplet of dark pions, $\pi_d$ expected to be the lightest. All heavier mesons have a lifetime of order $\Lambda_d^{-1}$ and decay to dark pions if kinematically allowed (i.e. $\rho \rightarrow \pi\pi$ in the SM). Dark baryons in these models are often very long lived. In the parameter regions we consider, dark pions are significantly lighter than dark baryons (as in QCD) and thus are typically produced in much larger abundances than dark baryons. 
This can be confirmed with SM data~\cite{Zyla:2020zbs} as well as a large $N_c$ expansion~\cite{Witten:1979kh} of QCD. Therefore we ignore the effects of dark baryons in this study, but of course these assumptions can be violated if the hadron spectrum is significantly different from that of the SM. 

For simplicity, we take a common mass of the dark pions, $m_{\pi_d}$, but for a study of non-trivial dark flavour dynamics, see~\cite{Renner:2018fhh}. The $Z'$ mediates a decay of the dark pions that can be computed using dark chiral perturbation theory with a width given by 
\begin{equation}\label{eqn:width}
\Gamma(\pi_d \rightarrow \bar{q}q) \approx \sum_q \frac{g_q^2 \,g_d^2 \,N_c \,m_q^2 \,f^2_{\pi_d}}{32\, \pi \,M^4}\, m_{\pi_d}
\end{equation}
where $N_c = 3$ is the number of SM colours, $f_{\pi_d}$ is the dark pion decay constant which we take to be $\sim\Lambda_d$, and $m_q$ is the mass of the SM quark in the final state. The sum over all SM quarks that are kinematically accessible, and we have ignored phase space effects, but they can be trivially added. The factor of $m_q^2$ is a spin-parity affect analogous to the decay of the charged pion in the SM, implying that the decay will be dominated by the heaviest quark kinematically accessible. We can estimate the proper decay length for a benchmark pion mass of 2 GeV:
\begin{equation} 
\begin{split}
c\tau_0 \approx 80 \,{\rm mm} \times \frac{1}{g_d^2\, g_q^2} \times 
	\left( \frac{2~{\rm GeV}}{f_{\pi_d}} \right)^2 \\
	\times \left( \frac{100~{\rm MeV}}{m_{ q}} \right)^2 
	 \left( \frac{2~{\rm GeV}}{m_{\pi_d}} \right)
	\left( \frac{M_{Z'}}{1~{\rm TeV}} \right)^4\,.
\end{split}
\label{eq:lifetime}
\end{equation}
which we see can be macroscopic but smaller than the size of an LHC detector for a wide range of parameter space.

Dark quarks (and thus dark jets) are produced at colliders like the LHC via an $s$-channel $Z'$. If kinematically accessible, resonant production where the $Z'$ is on shell will dominate the production. The lowest order cross section for this production process at a proton-proton collider is
\begin{multline}\label{eqn:pdfcross}
\sigma(pp \rightarrow Z' \rightarrow Q \bar{Q}) = \sum_{f = u,d} \int dx_{1}f_{f}(x_1)\int dx_{2}f_{\bar{f}}(x_2) \\ \times \frac{g_{d}^2 g_{q}^2}{72\pi} \bigg(\frac{  x_1 x_2  s}{ (x_1 x_2 s - M_{z'}^2)^2+ \Gamma^2 M_{Z'}^2} \bigg).
\end{multline}
Where $f_{i}(x_{i})$ is the parton distribution function for fermion $i$ and momentum fraction $x_{i}$. The Mandelstam variable $s$ is set to the centre of mass energy $\sqrt{s} = 13$ TeV. The total decay width of the $Z'$ is given by $\Gamma$, which has two contributions $Z'  \rightarrow q_i \bar{q}_i$ and $Z' \rightarrow Q_{i} \bar{Q}_i$,
\begin{equation}
\Gamma(Z' \rightarrow X\bar{X} ) \simeq 
\frac{N n g^2 M_{Z'}}{24 \pi}
\end{equation}
where for X being a SM (dark) quark, $N$ is the number of (dark) colours which we take to be 3, $n$ is the number of accessible flavours, and $g$ is the coupling to (dark) quarks. We have ignored kinematic factors which are only important if the $Z'$ is approximately degenerate with a pair of (dark) quarks. 

\begin{figure}[t]\label{fig:couplings}
\includegraphics[width=0.47\textwidth]{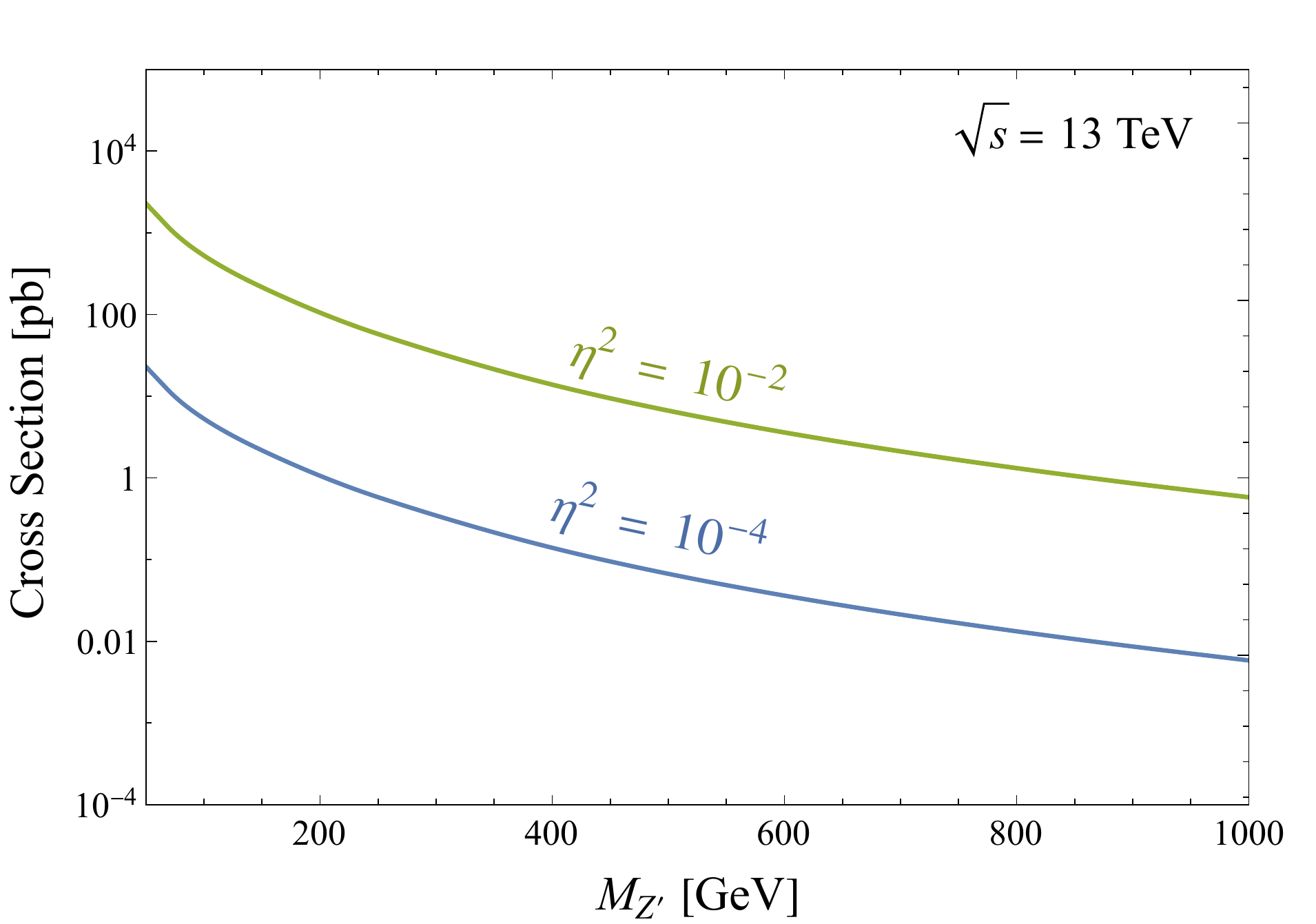} 
\caption{Cross section of $ pp \rightarrow Z' \rightarrow Q\bar{Q}$ for a leptophobic $Z'$ as a function of its mass $M_{Z'}$. Generated by convolving the partonic cross sections of $u,d$-quarks with their respective parton distribution functions at a centre of mass energy of $\sqrt{s} = 13$ TeV. $\eta$ is a function of the $Z'$ Lagrangian parameters defined in Eq.~(\ref{eqn:couplings}).}
\end{figure}

From these equations, assuming resonant production is dominant, we can show that to leading order the cross section for the process $ q\bar{q} \rightarrow Z' \rightarrow Q \bar{Q}$ depends only on the mass of the $Z'$ and the variable
\begin{equation}\label{eqn:couplings}
\eta^2 \equiv \frac{g_{d}^2 \cdot g_{q}^2 }{g_{d}^2 +\big( \frac{n_{f}}{n_{d} }\big) g_{q}^2}, 
\end{equation}
scaling with $\eta^2$. We have assumed the number of dark colours is also 3. The production cross section at a centre of mass energy $\sqrt{s} = 13$ TeV is shown in Fig.~\ref{fig:couplings} as a function of mass for a couple of benchmark values of $\eta^2$. We see that number of such events at the LHC with an integrated luminosity $\sim 100 - 3,000$ fb$^{-1}$ can be very large. 

\begin{figure*}\label{fig:feynman}
\centering 
\includegraphics[scale=.85]{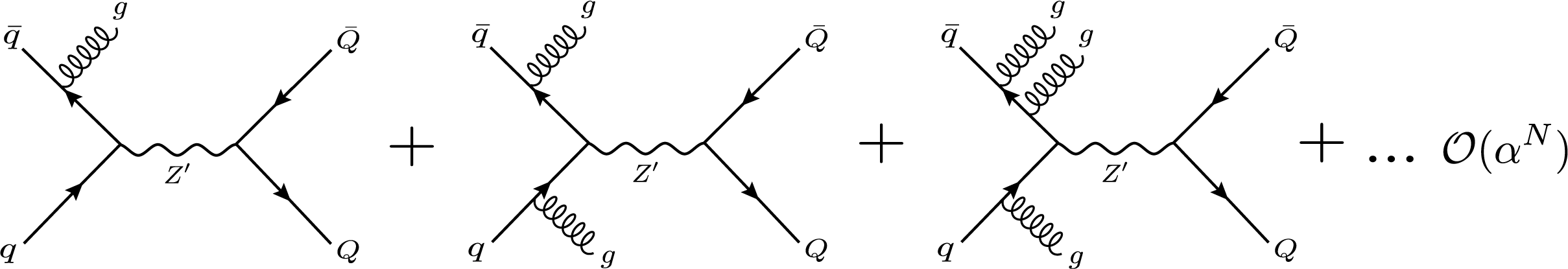} 
\caption{Higher order QCD initial state radiation diagrams of $q\bar{q}$ induced $Z'$ resonant production. Gluons are radiated from the quark lines in the form of detectable hard QCD jets. EW splittings are also possible with diagrams scaling with the EW couplings. }
\end{figure*}

Alongside the LO contribution to the $Z'$ production are higher order terms from initial state radiation (ISR) coming off of the incoming quark lines. In Fig.~\ref{fig:feynman} we show this for QCD gluon radiation. These gluons will hadronize into additional hard QCD jets affecting the overall event's topology. ISR is not exclusive to QCD, the quarks may radiate a hard $W$, $Z$ or photon. These ISR contributions have a smaller rate than the leading order process, but they can be easier to detect experimentally. 

In the scalar mediator case explored in~\citep{Schwaller:2015gea}, collider and direct detection experiments put on stringent constraints on hidden sector spectrum~\citep{Mies:2020mzw}. In our case, these $Z'$ models are of large interest for detector searches because of the freedom in the production cross section of dark hadrons. This model, however, does have resonant dijet production proportional to $g_{q}^4$ which can place constraints~\citep{PhysRevD.96.052004,2019316,Sirunyan_2019}. Most of these studies at modern collider experiments are only sensitive to the higher mass regime $M_{Z'} > 1$ TeV because of the filtering of lower mass events by their triggers. Some searches have employed data scouting techniques and trigger level analysis to probe lower masses ~\cite{2017520,PhysRevLett.121.081801}. Typical upper bounds at the lower mass regime are $g_{d} < 0.1$. We do demand that the dark quarks produce emerging jets with lifetimes between [1mm - 1000m]. Doing so puts constraints on the product of couplings $g_{d} \cdot g_{q}$ from Eq.~(\ref{eqn:lagrangian}). Assuming lower mediator masses, dark pion masses, and decay constant at $\mathcal{O(\textrm{GeV})}$ emerging jet events are consistent with $g_{d} \cdot g_{q} \lesssim 0.2$. 

For the rest of this work, we take $c\tau$ to be a free parameter. The mass of the $Z'$ and $c\tau$ are related by Eq.~(\ref{eq:lifetime}), but there are enough parameters in the models such that we can tune each variable independently. Therefore we vary the mass of the $Z'$ as it broadly controls the total energy in the event. 
For the dark QCD parameters, we use six benchmark parameter points shown in Table \ref{tab:benchmarks}. The first two, Model $\textbf{A}$ and Model $\textbf{B}$, have different dark pion masses of 2 GeV and 5 GeV and 2 different lifetimes of 5 mm and 150 mm, respectively. These models are used in studying current triggers in Section~\ref{sec:current}.\footnote{These were also the benchmarks used in~\cite{Schwaller:2015gea}.} Models $\textbf{C}$, $\textbf{D}$ and $\textbf{E}$, are identical to Model $\textbf{A}$ but with lifetimes ranging from 5mm to 500 mm. Model $\textbf{F}$ has the same lifetime as Model $\textbf{D}$ but a heavier hadron spectrum. 
Models $\textbf{A}$, $\textbf{C}$, $\textbf{D}$, $\textbf{E}$ and $\textbf{F}$ are considered in the machine learning trigger analysis of Section~\ref{sec:ML}. 
  
\begin{table}[b]
\centering
\begin{tabular}{ccccccc}
\hline
Model & \textbf{A} & \textbf{B} &  \textbf{C} & \textbf{D} & \textbf{E} & \textbf{F} \\ \hline 
$\Lambda_d$ & 10 GeV  & 4 GeV & 10 GeV & 10 GeV & 10 GeV & 20 GeV  \\
$m_V$ & 20 GeV  & 8 GeV & 20 GeV & 20 GeV  & 20 GeV & 40 GeV \\
$m_{\pi_d}$ & 5 GeV  & 2 GeV & 5 GeV & 5 GeV & 5 GeV & 10 GeV \\
$c\,\tau_{\pi_d}$  & 150 mm  & 5 mm & 50 mm & 500 mm & 5 mm & 500 mm\\ \hline
\end{tabular}
\caption{Dark sector parameters for our benchmark models. $\Lambda_d$ is the dark confinement scale, $m_V$ is the mass of the dark vector mesons, and $m_{\pi_d}$ is the pseudo-scalar mass. $c\,\tau_{\pi_d}$ is the rest frame decay length of the pseudo-scalars. We take $N_c = 3$ and $n_f = 7$ in both benchmarks. Models \textbf{A} and \textbf{B} are considered in Section~\ref{sec:current}, while Models \textbf{A}, \textbf{C}, \textbf{D}, \textbf{E} and \textbf{F} are considered in Section~\ref{sec:ML}.}
\label{tab:benchmarks}
\end{table}

\section{Event generation}
\label{sec:event}

Here we describe our simulation pipeline to generate Monte Carlo events for the models described in the previous section. The events were generated using a modified spin-1 mediator model\footnote{https://github.com/smsharma/SemivisibleJets}~\cite{Cohen_2017} implemented using the \verb!FeynRules!~\cite{Alloul_2014} package. The model is outputted as a UFO~\cite{Degrande_2012} file which allows generation of hard processes with \verb!Madgraph5_aMC@NLO!~\citep{Alwall:2014hca}, and we use LHC conditions with a centre of mass energy of 13 TeV. This output is interfaced to the Hidden Valley~\cite{Carloni_2010,Carloni_2011} module of \verb!Pythia8!~\citep{Sjostrand:2014zea}, which simulates showering and hadronization in the dark sector as well as decays of dark hadrons to either other dark hadrons or to SM states. Initial state radiation (ISR) in QCD or EW, i.e jets, leptons and EW gauge bosons, are included in the hard processes in \verb!Madgraph5_aMC@NLO!, and we use \verb!Pythia8! to shower and hadronize SM quarks.

Double counting of jets can occur when introducing ISR at the matrix element (ME) level and then subsequently hadronizing the partons in \verb!Pythia8!. To avoid this, we use the MLM matching and merging procedure~\cite{Mangano_2007}. An \verb!XQCut! of $M_{Z'}$/10 is used at the matrix element level which forces the production of only partons with a minimum $K_{T}$ separation. Matching in \verb!Pythia8! is done by enforcing $ \verb!QCut! > \verb!XQCut!$. For photon ISR, a minimum transverse momentum cut is placed on additional photon radiation of $P_{T} > 10 \text{ GeV}$ to avoid soft and collinear divergences. The resulting hadrons are clustered into jets using the Anti-$k_{t}$ algorithm~\cite{Cacciari_2008} implemented in \verb!FASTJET!~\citep{Cacciari:2011ma}, where the ATLAS tracker's pseduorapidty is $|\eta| < 2.49$ and the jet angular parameter $R = 0.4$. 

For the analysis of current triggers in Section~\ref{sec:current}, one million events are generated for each $Z'$ mass within the range [50 GeV, 1500 GeV], with a step size of 50 GeV, for Models \textbf{A} and \textbf{B} in Table~\ref{tab:benchmarks}. A $Z'$ width of $\Gamma_{Z'} = M_{Z'}/100$ is used throughout. A crude detector volume cut is implemented at the \verb!Pythia8! stage. All particles that have not decayed outside of a cylinder of $(r = 3000$ mm$, z = 3000$ mm) are considered stable. This cut was placed to mimic the detector volume's reach up to the muon spectrometer.
For the analysis of current triggers, backgrounds rates are already known and do not have to be estimated.

In Section~\ref{sec:ML}, the focus is on using a machine learning approach for novel triggers. We are less interested in hard ISR events and therefore use \verb!Pythia8!'s hidden valley production process $f\bar{f} \rightarrow Z_{v}$ processes instead of \verb!Madgraph5_aMC@NLO!. Hits on the ATLAS inner tracker are used as discriminating variables. A proper detector simulation of the inner tracker is outside of the scope of this work, but we use a crude detector simulations with code used in~\citep{Knapen_2017} which encompasses the ATLAS tracker from the Inner Bilayer (IBL) to the Transition Radiation Tracker (TRT). This detector simulation assumes simple models of energy loss through each thin layer of the detector. An emerging jet registers various hits as function of the radial distance from the interaction point. Section~\ref{sec:ML} considers Models \textbf{A}, \textbf{C}, \textbf{D}, \textbf{E} and \textbf{F}, and we choose a $Z'$ mass of $M_{Z'} = 500$ GeV. This set of models span a wide range of lifetimes while keeping all other model parameters equal. 

When considering new triggers, we must also estimate the background. Various background processes are considered, but the backgrounds are dominated by $ pp \rightarrow b\bar{b}$ events which have high multiplicity and hadrons with longer lifetimes producing many displaced vertices that can mimic an emerging jet signal. We simulate $10^5$ events of $ gg \rightarrow b\bar{b}$ using \verb!Pythia8!'s heavy flavour hard $b\bar{b}$ processes. Pileup added to both signal and background events with \verb!Pythia8!'s minimum bias events.  For each signal or background event, a number of minimum bias events are added randomly sampled from a poisson distribution with mean of $\mu =  50$, approximately corresponding to expected pileup contributions for recent and near future ATLAS runs. The inclusive background cross section $\sigma_{\text{bkg}}$ is taken from the \verb!Pythia8! event generation stage, which is used to estimate the instantaneous background rate.

\section{Current triggers}
\label{sec:current}

A consequence of having high instantaneous luminosities, such as at the LHC, is the necessity of implementing triggering streams on specified event criteria. ATLAS/CMS produce event rates in the MHz range, which is far too large to write every event onto record. Triggers were introduced to greatly reduce the event rate that is written for offline use, to a manageable $\sim$ 1 kHz. Although emerging jet experimental searches do exist~\citep{Sirunyan:2018njd}, models with unique phenomenology such as emerging jets may not be visible to the current dedicated trigger sets used at ATLAS and CMS~\citep{Owen:2302730,Khachatryan_2017}. In this section, we quantify the efficiency of current triggers for emerging jet phenomenology.
 
In addition to leading order production, we also study the effects of ISR on trigger efficiencies which come from Feynman diagrams of the type shown in Fig.~\ref{fig:feynman}. Multiple QCD jets (and/or electroweak gauge bosons) can modify the naive expectation of a two-pronged emerging jet scenario. Each additional hard object will change the event's topology from back to back scattering in the transverse plane. Emerging jets on their own provide unique detector signals that may have difficulty passing trigger selections in various parameter spaces. We will show that the inclusion of hard SM objects, such as jets and leptons, increases the likelihood of passing the triggers. Numerical trigger thresholds used in ATLAS~\cite{ATL-DAQ-PUB-2018-002} and used for this analysis are shown in Table~\ref{tab:triggers}.

\begin{table}[b]
\centering
\begin{tabular}{cccc}
\hline
Triggers & Lower (GeV) & Higher (GeV) & Offline (GeV)  \\ \hline 

Single Jet & 100 & 420 & 435  \\
Single $\gamma$ & 20* & 140 & 145 \\ 
Single e  & 22* & 26* & 27  \\ 
Single $\mu$ & 20 & 26* & 27 \\ 
MET & 50 & 110 & 200  \\
$H_T$ & // & 850 & //\\
\hline
\end{tabular}
\caption{The 2017 ATLAS triggers~\cite{ATL-DAQ-PUB-2018-002} used in Section~\ref{sec:current} analysis. Triggers are separated by lower level (L1) thresholds, Higher level (HLT) thresholds and offline selection criteria. Reconstructed jets used in the Single Jet, MET and $H_{T}$ trigger have R = 0.4. Thresholds with (*) must additionally satisfy the isolation cone criteria in Eq.~(\ref{eqn:Icone}). A lower level threshold isn't given for the $H_{T}$ as we seed it from the lower level Single Jet trigger instead. }
\label{tab:triggers}
\end{table}

\subsection{Description of triggers}

\begin{figure}
\centering
\includegraphics[scale=.5]{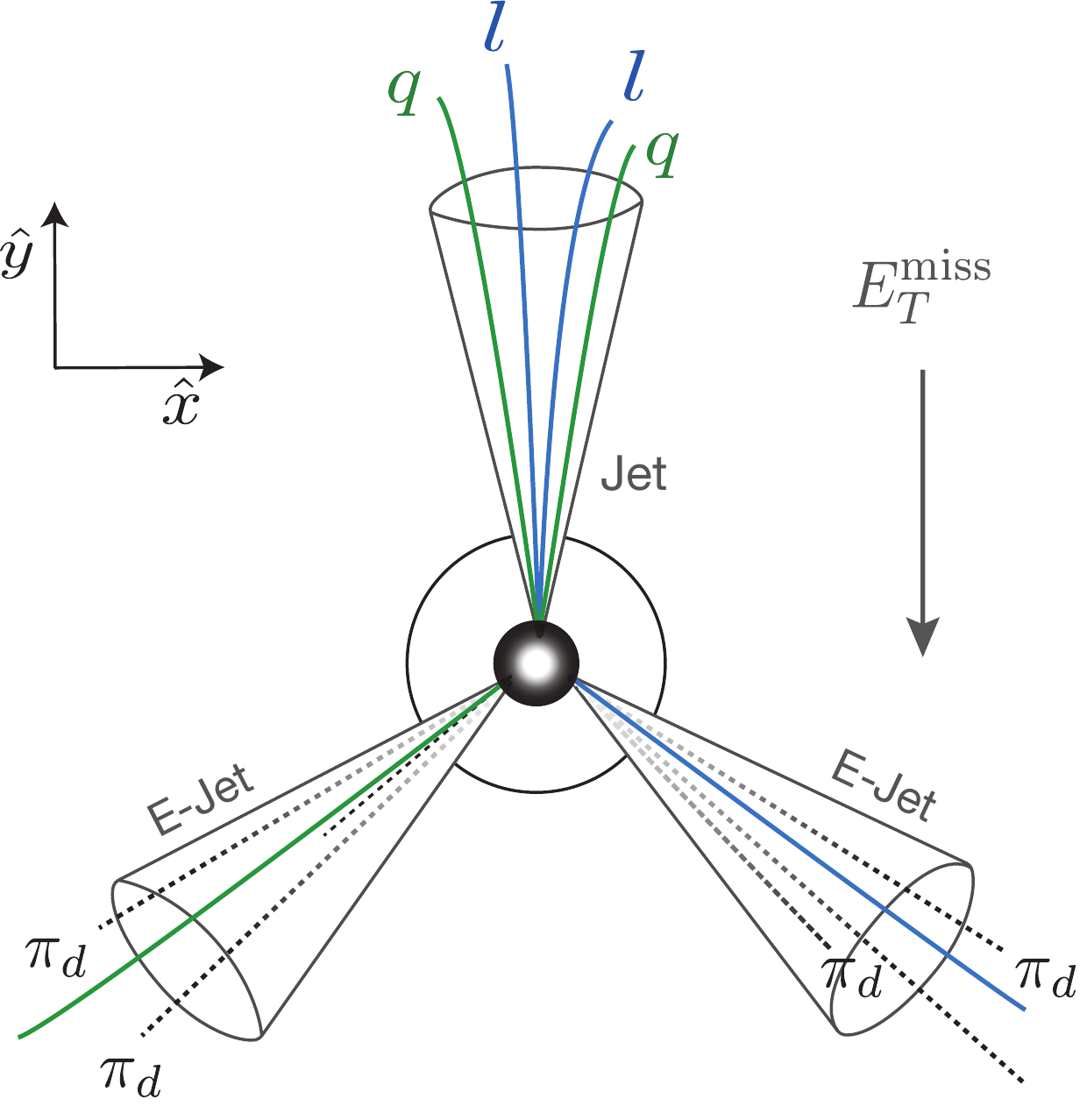} 
\caption{The Mercedes topology of three jets recoiling off of each other within the $p_{T}$ plane. In this case, two emerging jets carrying missing transverse energy recoil off of a visible hard QCD jet. This configuration produces more missing transverse energy as the jet momentums are balanced in opposite directions. }
\label{fig:mercedes}
\end{figure}

\begin{figure*}
\centering
\begin{minipage}[c]{\textwidth}
\includegraphics[width=.32\textwidth]{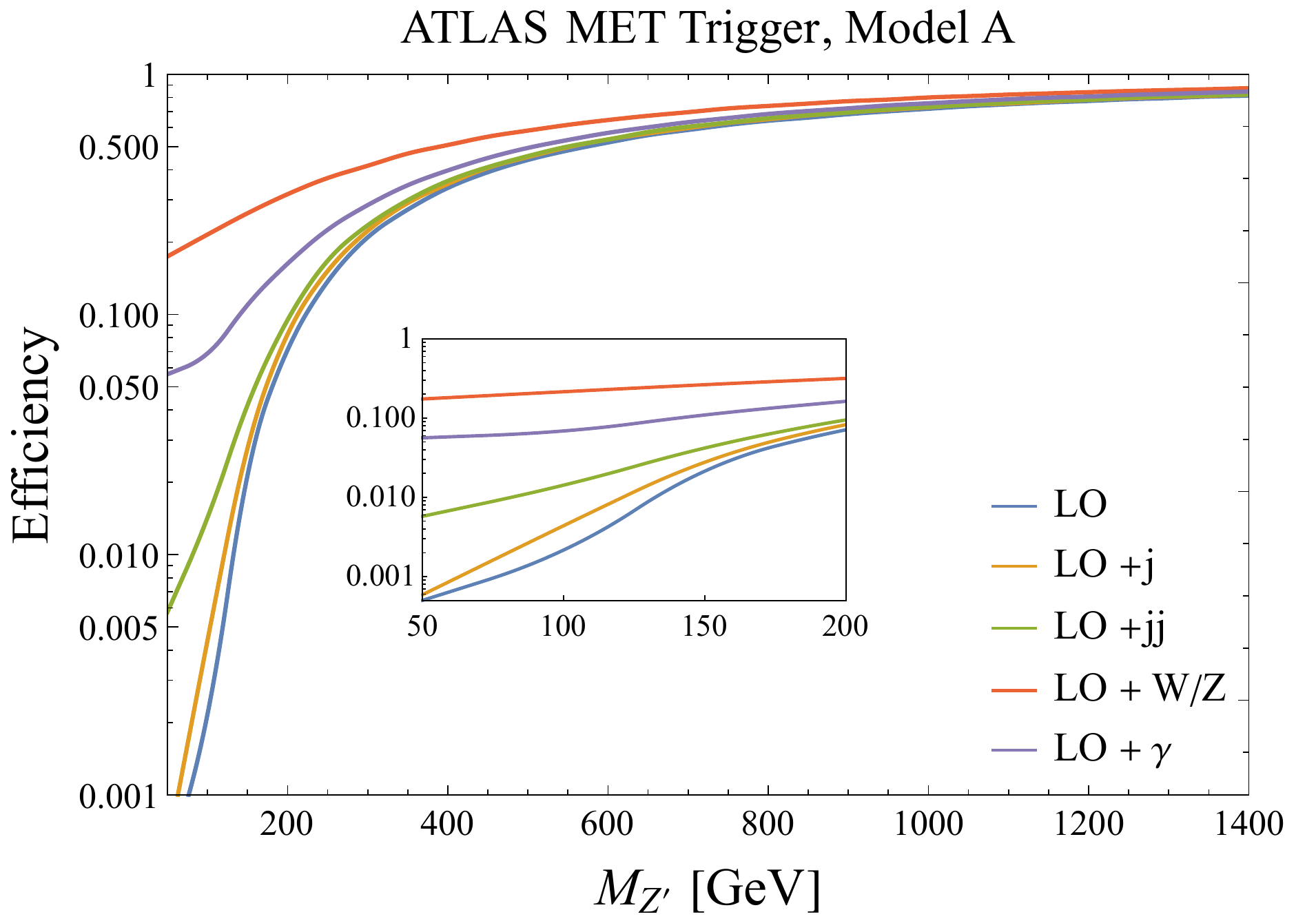} 
\hfill
\includegraphics[width=.32\textwidth]{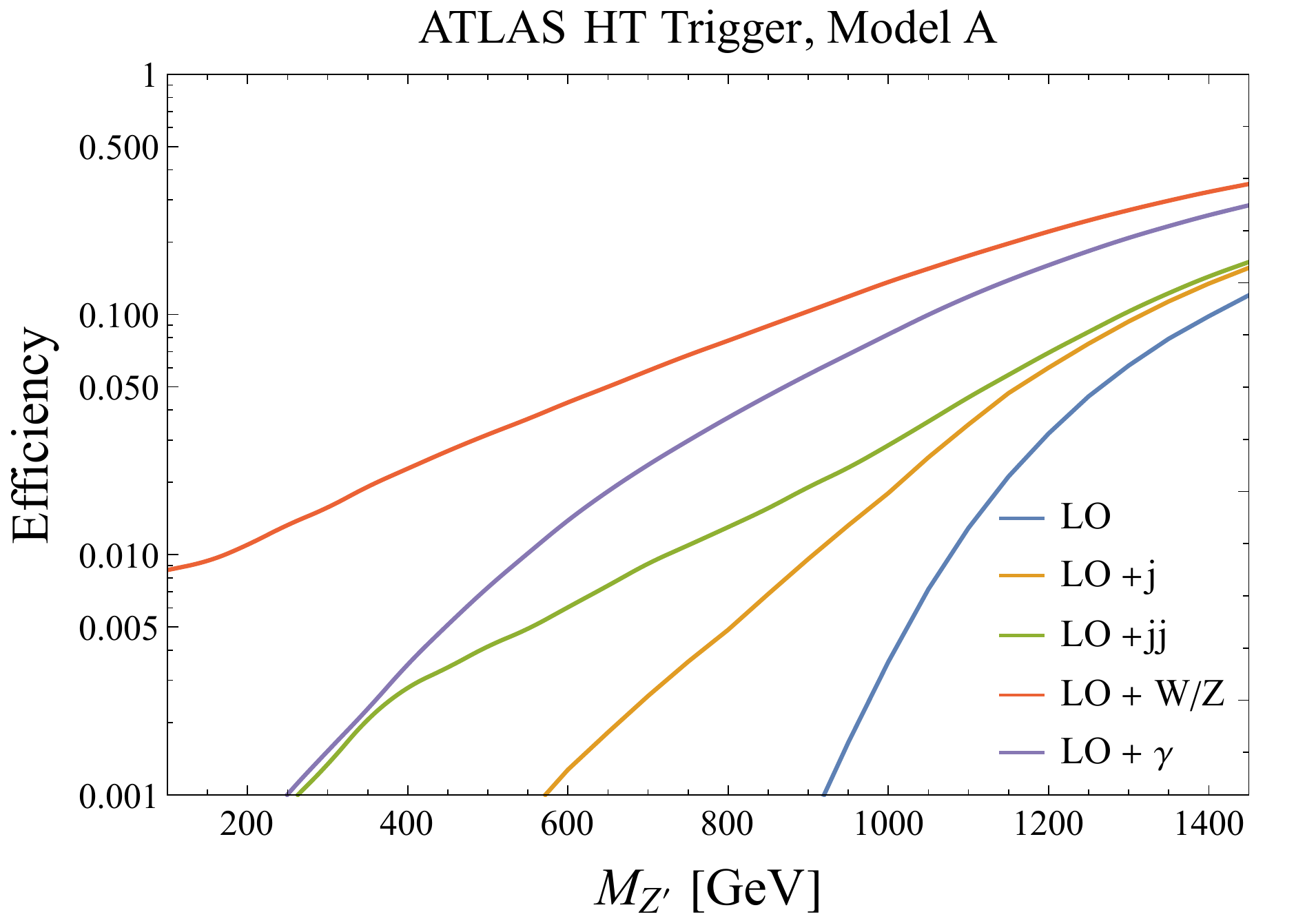} 
\hfill
\includegraphics[width=.32\textwidth ]{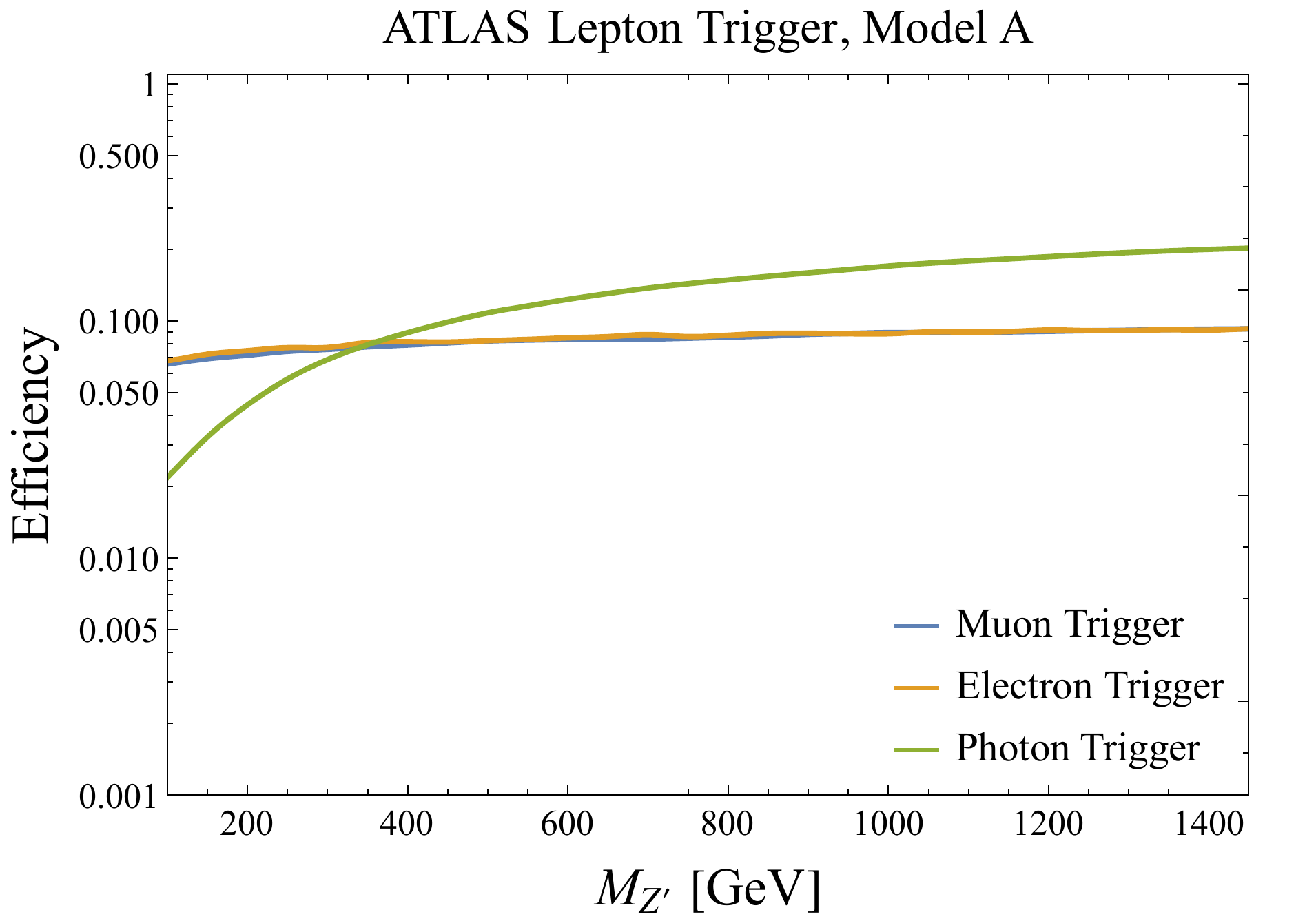}
\hfill
\includegraphics[width=.32\textwidth]{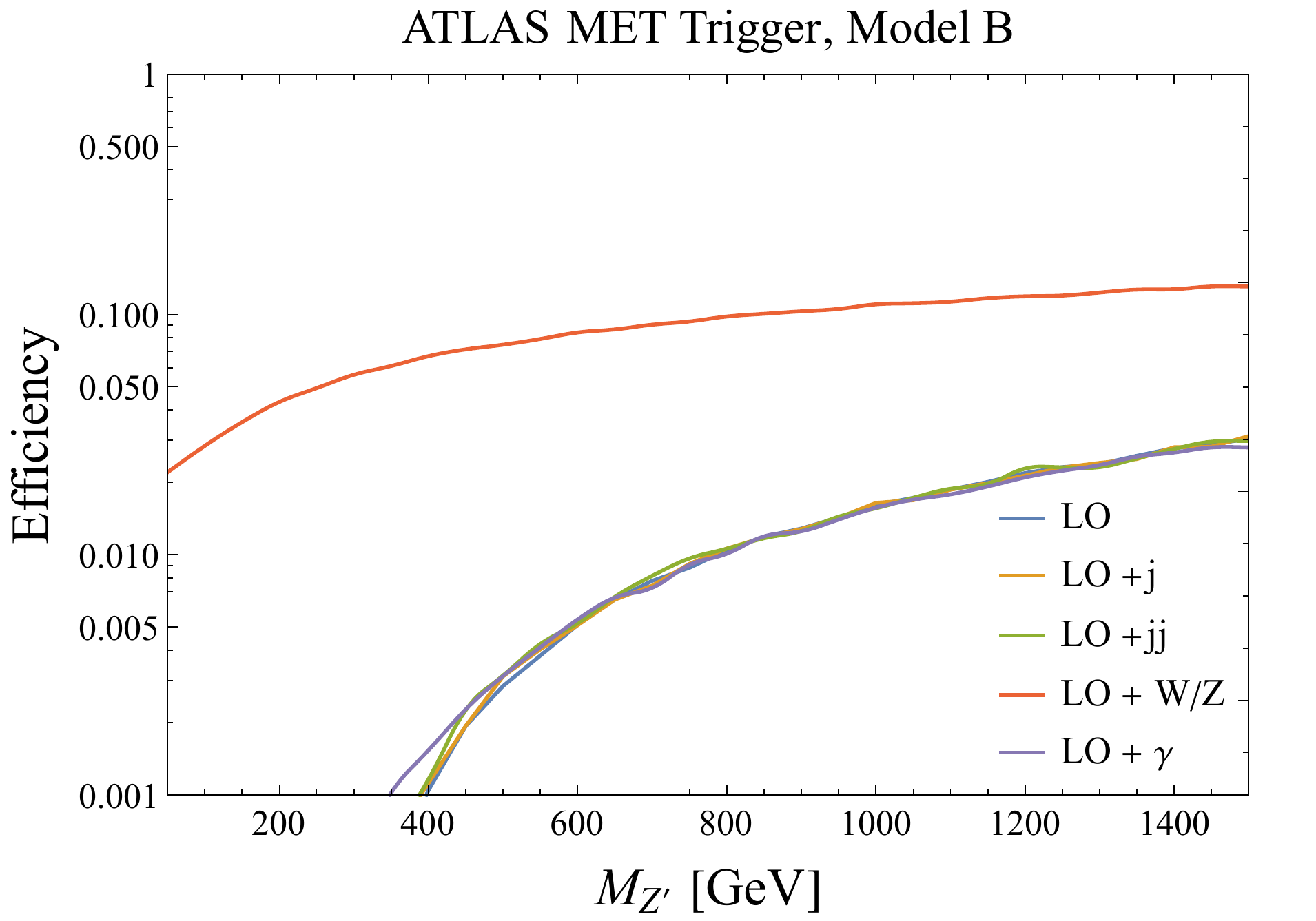} 
\hfill
\includegraphics[width=.32\textwidth]{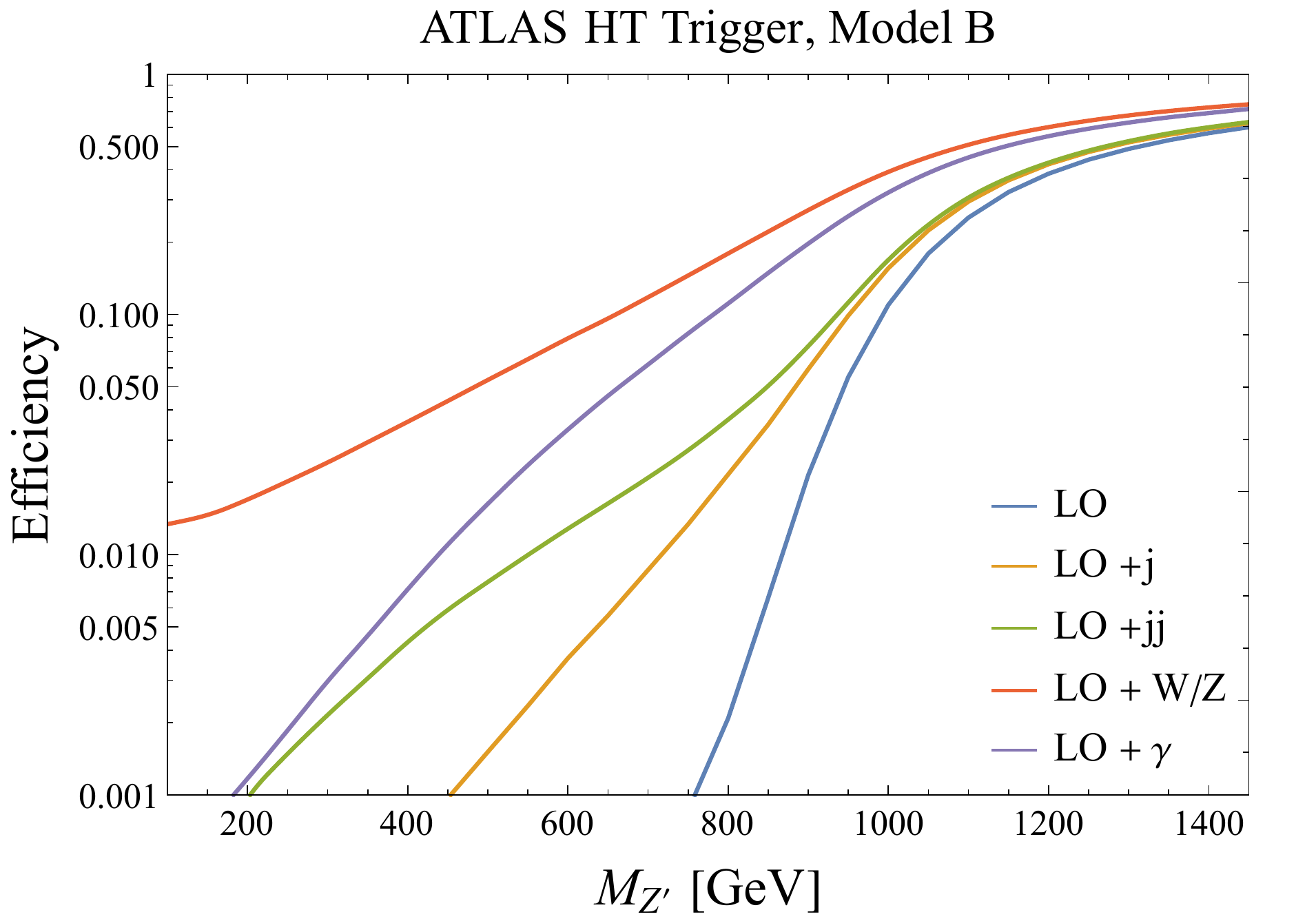} 
\hfill
\includegraphics[width=.32\textwidth]{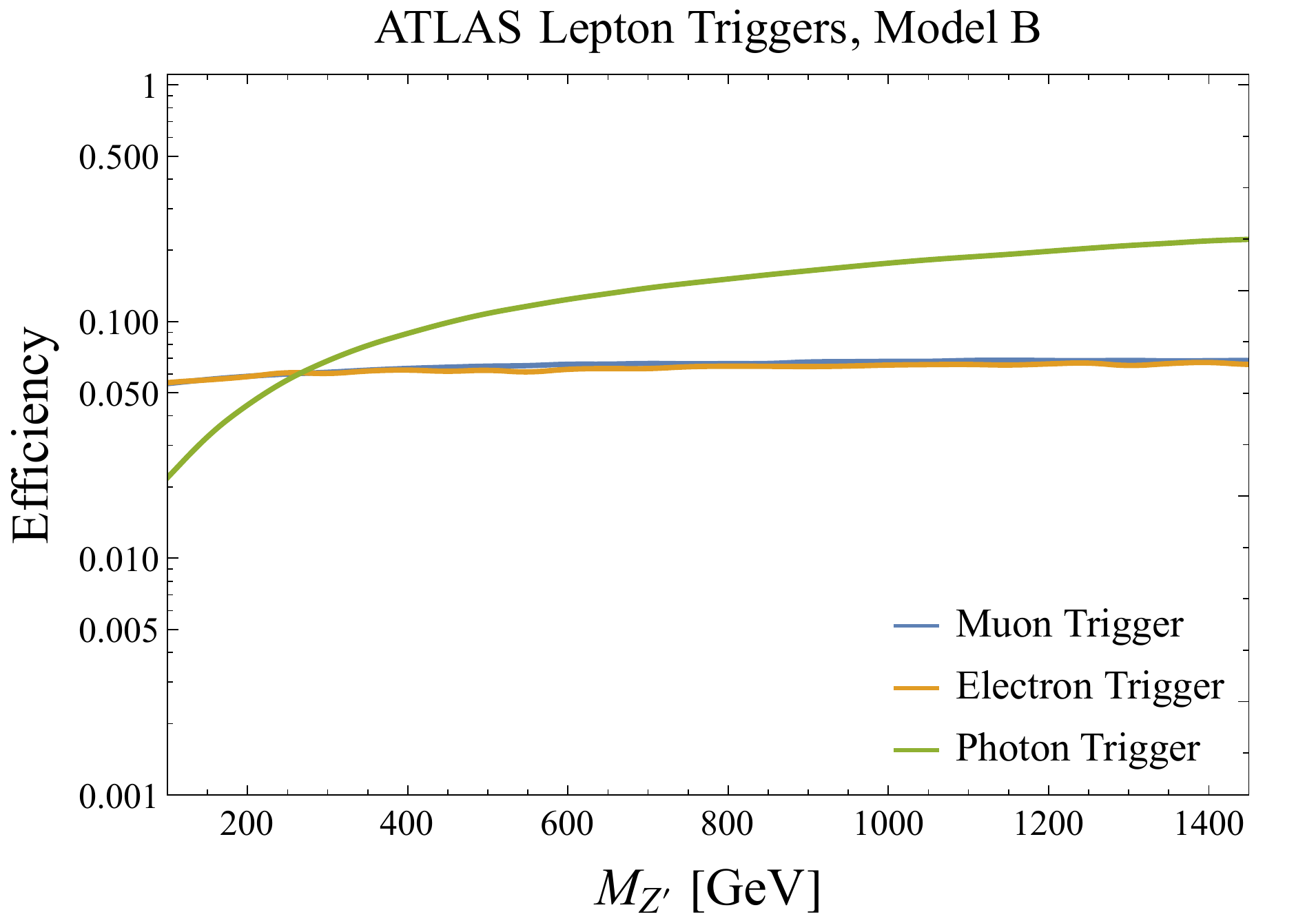} 
\end{minipage}
\hfill
\caption{ATLAS trigger efficiencies, at truth level, for various processes outlined in Section~\ref{sec:event}. The first (second) row corresponds to Model \textbf{A} (\textbf{B}) from Table~\ref{tab:benchmarks}. In the first column, we have the MET triggers; in the second column, the $H_{T}$ triggers; and, in the third column, the lepton and photon triggers. Both the electron and muon trigger efficiencies considered $W/Z$ ISR while the photon trigger considered photon ISR only. All processes were generated under LHC conditions of a centre of mass energy of 13 TeV. The trigger thresholds were taken from the ATLAS trigger menu summarized in Table~\ref{tab:triggers}. 
}
\label{fig:eff}
\end{figure*}
The trigger systems at both experiments are separated into two disjoint online subsystems: the low level hardware trigger system (L1), and the high level software trigger system (HLT). L1 primarily deals with low level information from energy depositions in the calorimeter towers and minimally reconstructed jet variables. Track reconstruction and jet algorithms are available at HLT triggers for more sophisticated triggering criteria. Here, we will describe the triggers that are relevant to our analysis. 

\underline{\textit{MET Triggers}}: Dark sector mesons with sufficiently long lifetimes $\tau_{d}$ will typically escape the detector before decaying and thus contribute to Missing Transverse Energy (MET). Energy deposited within the calorimeter towers is reconstructed for MET calculations at both ATLAS and CMS. In the plane transverse to the beam, the transverse momentum $P_{T}$ is conserved with a zero net $P_{T}$, and thus this observable is sensitive to production of invisible particles. Energy deposition in each tower is summed and a transverse momentum vector is constructed using
\begin{equation}\label{eqn:MET}
P_{T} = \sqrt{ \big(\sum P_{x}\big)^2 + \big(\sum P_{y}\big)^2}.
\end{equation}
Any non-zero contribution is taken to be MET $= |P_{T}|$. 

Triggers cut on an event's MET according to the trigger menu thresholds in Table~\ref{tab:triggers}, at both L1 and HLT. L1 thresholds are lower than the HLT, although the calculated MET might differ between reconstructed calorimeter tower energies at both levels. It is usually precise enough to assume that HLT is seeded from L1 with $100\%$ efficiency, therefore we only consider the HLT efficiencies. 

Typical $Z'$ events, even if the dark hadrons have long lifetimes, tend to have relatively low MET because the two emerging jets are produced back to back, so there can be significant cancellation between them. Hard initial state radiation can qualitatively change this picture as shown in Fig.~\ref{fig:mercedes}. With additional radiation the two emerging jets are no longer collinear, and their MET will to some extent add. Furthermore, the additional radiation means that the energy of each jet will be larger (for fixed $Z'$ mass), which also tends to increase the MET. Fig.~\ref{fig:mercedes} shows QCD radiation, but the same logic applies to EW radiation. 

\underline{\textit{$H_T $Triggers}}: The $\textit{H}_T$ trigger is a threshold on the scalar $P_{T}$ sum of the event's reconstructed objects. These triggers help reduce rates by focusing on events with large final state transverse energy $\textit{E}_T$. The $\textit{H}_T$ is built from objects with $|\eta| < 2.5 $ and jets are only included if they have $\textit{P}_T \ge 50\textrm{ GeV}$. 

For the $Z'$ model here, as the mediator mass increases, the trigger is more likely to be satisfied due to the larger final state momentum imparted onto the constituents. 
At lower mediator masses, the efficiency can be increased with additional hard QCD radiation.
The QCD jets are more visible than the dark sector jets and lead to a higher $\textit{H}_T$. This is also realized in the EW ISR case, as the hard photon/$W$/$Z$ caries all of the ISR energy,\footnote{In leptonic decays of the $W$ or decays of the $Z$ to neutrinos, the energy of the neutrino does not contribute to $H_T$. } and if properly reconstructed, contributes significantly to final state $\textit{E}_T$. This effect can be seen in Fig.~\ref{fig:mercedes}.

\underline{\textit{Lepton and Photon Triggers}}: ISR of a $W^{\pm}, Z$ can add additional hard leptons and/or missing energy with neutrinos, and radiation of photons can also be used. Triggers that cut on identified leptons ($e$ and $\mu$) and photons, are considered. The trigger menu hosts a range of triggers depending on the level of reconstruction necessary for the identification of the event's particles. 

The rate of leptons produced within jets is extremely high, so lepton triggers also have isolation requirements. 
 ATLAS identifies leptons under various classifications. These criteria are known as tight (loose) Isolation~\cite{Aad_2020,Aaboud_2019}, which are defined by
\begin{equation}\label{eqn:Icone}
 \sum_{i  \;\in \;\textrm{cone}} \frac{ P_{T}^i(\Delta R < R^{l})}{P_{T}^l} < \mathcal{I},
\end{equation}
where in Eq.~(\ref{eqn:Icone}), $\mathcal{I} = 0.6$  $(1.5)$ for tight (loose) isolation, $P_{T}^l$ is the transverse momentum of the candidate lepton, $P_{T}^l$ are the transverse momentum of the visible non-candidate objects within the isolation cone, $\Delta R$ is the distance between the $i^{th}$ particle and the candidate lepton $l$, and $R^{l} = 0.2$ $(0.3)$ is the cone radius for electrons (muons). This is accomplished with $P_{T}$ calculated from lepton track measurements. 

In the case of photons, the condition is given by
\begin{equation}\label{eqn:IconeGamma}
 \sum_{i  \;\in \;\textrm{cone}} E_{T}^i (\Delta R < R^{\gamma}) < 0.022 \cdot E_{T}^{\gamma} + 2.45 \text{ GeV}.
\end{equation}
The photon isolation uses calorimeter measurements of the transverse energy $E_{T}$ since photons do not leave tracks, $E_{T}^{\gamma}$ is the transverse energy of the candidate photon, and $E_{T}^i$ of the $i^{th}$ cone constituent. Since we are simulating events without full detector effects, we assume that the truth level transverse energy of the photon is equal to that of the reconstructed calorimeter energies. 

Trigger menus may demand different levels of isolation strictness between L1 and HLT. In Table~\ref{tab:triggers}, for the single electron trigger considered, both L1 and HLT must adhere to tight isolation criteria where as the muon trigger has isolation only at L1. It is important to consider these drastic differences of kinematic acceptance between L1 and HLT 
when calculating the total efficiency. Because of this, we do not assume that the lepton triggers are seeded from an L1 trigger with $\epsilon_{L1} = 1$ (100$\%$ efficiency). Instead, we calculate the L1 efficiency and project the product $\epsilon = \epsilon_{\text{L1}} \cdot \epsilon_{\text{HLT}}$ in our results.

\subsection{Results with current triggers}

We first calculate the trigger efficiency for different triggers in Model \textbf{A} and Model \textbf{B}. In this section we do not use a detector simulation as the output of \verb!Pythia! should be a reasonable estimate of these simple variables. The efficiency is the number of events that pass the threshold, and therefore get written for offline use, over the total number of events, and these are shown in Fig.~\ref{fig:eff}, with the top (bottom) row being for Model \textbf{A} (\textbf{B}). 
In the first column we see that for Model \textbf{A}, any kind of radiation increases the MET trigger efficiency because in Model \textbf{A} the dark hadrons have long lifetimes, and initial state radiation increases their momentum and makes them not back to back. Radiation of $W/Z$ does the best because of the presence of neutrinos. For Model \textbf{B}, notice that we only get significant improvement with $W/Z$ radiation. 

In the second column of Fig.~\ref{fig:eff} we see that QCD radiation can significantly increase the efficiency of the $H_{T}$ trigger for both lifetime benchmarks, and two hard jets does better than a single extra jet. Both $W/Z$ and photon radiation do better than QCD radiation because of the the clean visible momentum carried by the EW radiation. At low masses $M_{Z'} \lesssim 500 \text{ GeV}$, the efficiencies are similar for both models while at higher masses $M_{Z'}\gtrsim  500 \text{ GeV}$, Model \textbf{B} becomes far more efficient as it carries more visible particles in the final state. The improvement due to extra radiation is very important at low mass, a section of the $Z'$ parameter space not easily probed, but less so at high mass because the trigger can already be quite efficient at leading order in that case. 

\begin{figure}[t]
\centering 
\includegraphics[width=.48\textwidth ]{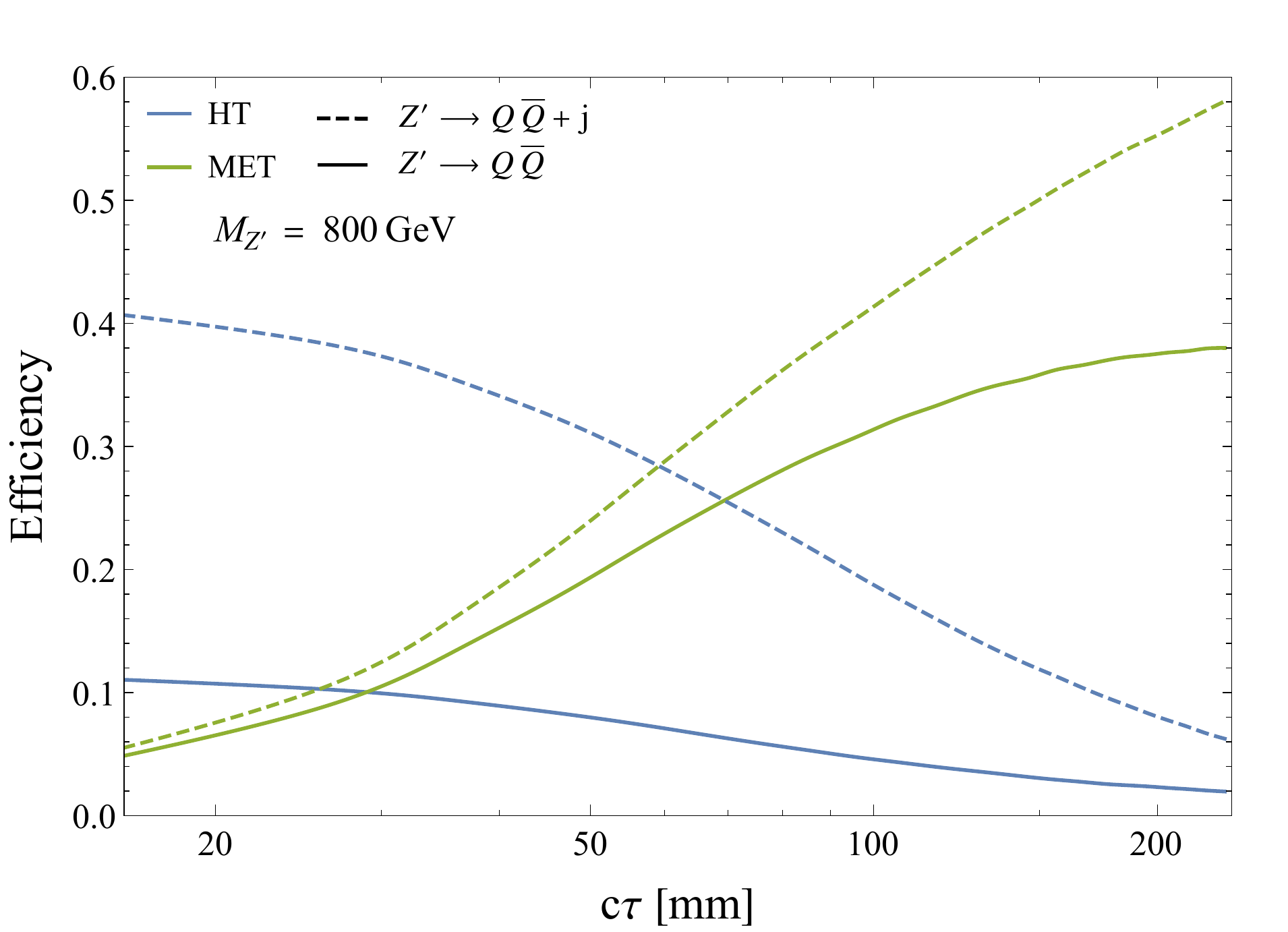}
\caption{ATLAS trigger efficiencies as a function of the dark pion's lifetime $c\tau$. The LO process is shown in the solid lines, the LO + j process is shown in the dashed lines. A inverse relationship is exhibited between MET and $H_T$ efficiencies.  }
\label{fig:efflifetime}
\end{figure}

We also show the trigger efficiency as a function of dark pion lifetime for a fixed $Z'$ mass of 800 GeV in Fig.~\ref{fig:efflifetime}. As expected, as the lifetime increases, more of the energy escapes the detector and the MET trigger gets better while the $H_{T}$ trigger gets worse. 
This explains the differences of the first and second columns of Fig.~\ref{fig:eff} between both Models \textbf{A} and \textbf{B}.

\begin{figure*}
\centering
\begin{minipage}[c]{\textwidth}
\includegraphics[width=0.48\textwidth]{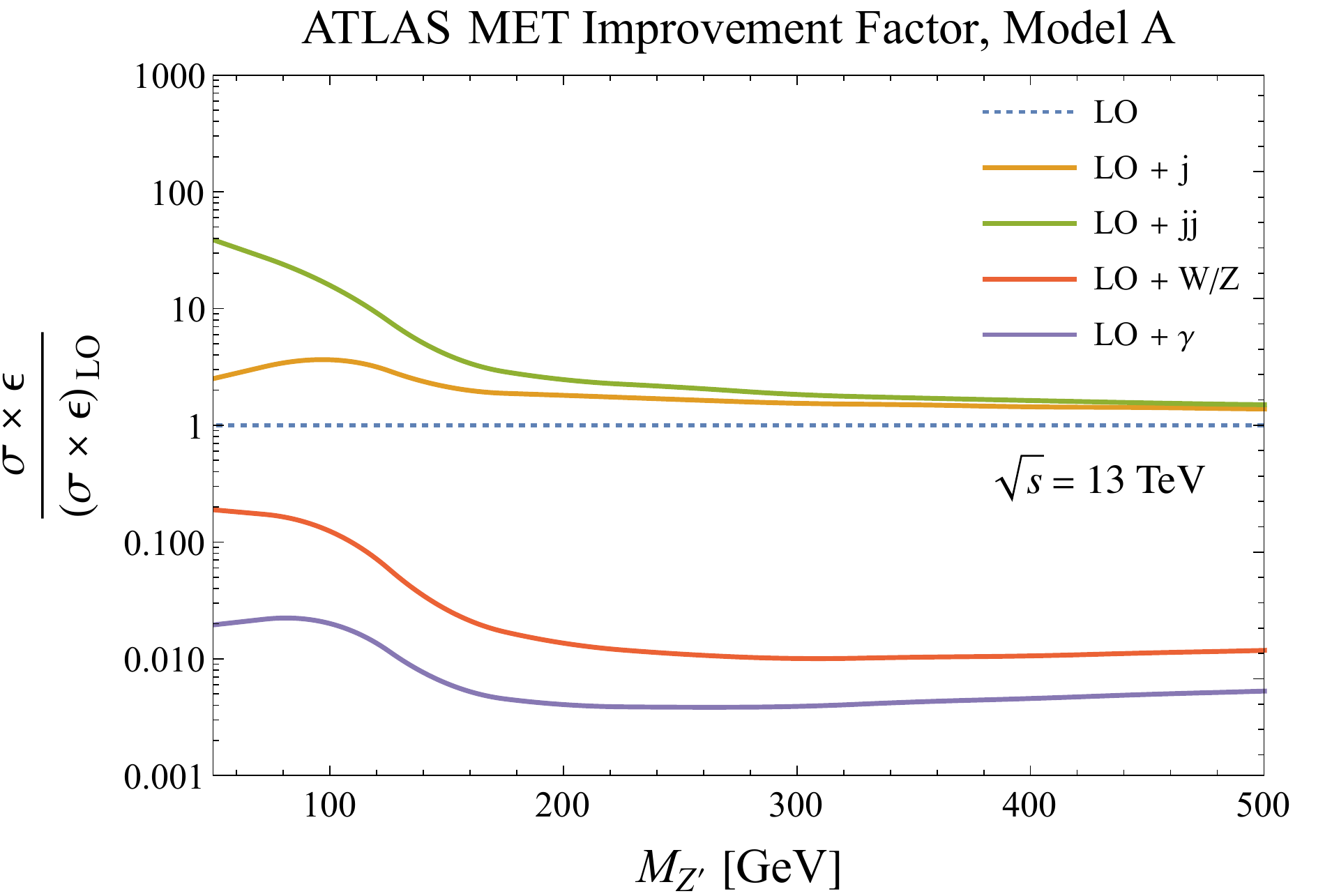}
\hfill
\includegraphics[width=0.48\textwidth]{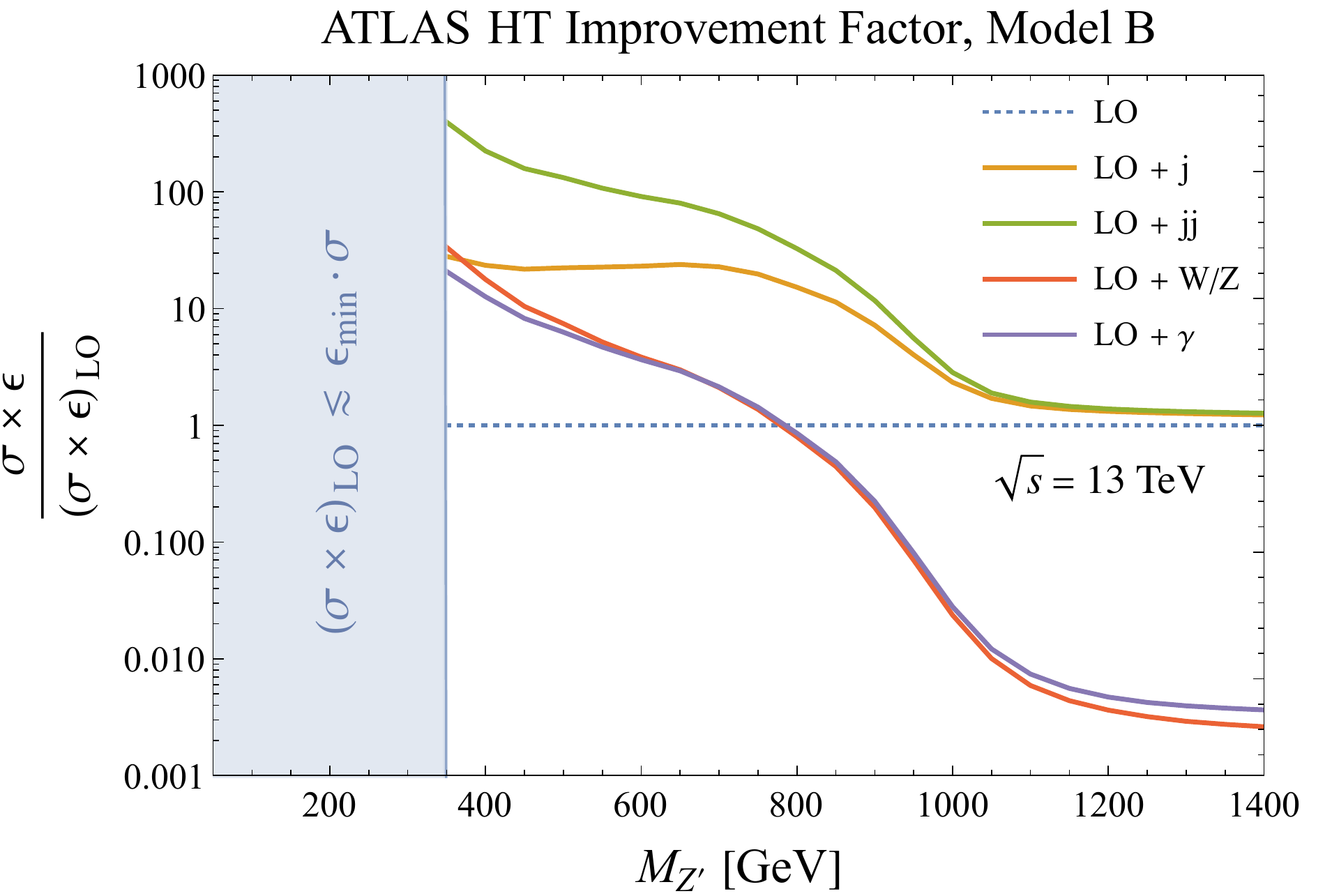}
\end{minipage}
\hfill
\caption{Cross section times efficiency of various processes (leading order, 1-jet ISR, 2-jet ISR, electro weak ISR) scaled by their respective leading order process. The left plot uses the MET trigger and Model \textbf{A}, while the right uses $H_{T}$ and Model \textbf{B}. The dotted line is the leading order process, which when normalized to itself, is just unit as a reference for the rate boost obtained from additional processes. On the right plot, the blue region on the left has zero events that we simulated pass the trigger, $\epsilon \lesssim \epsilon_{\rm{ min}} = 1/ 400$ $000$. }
\label{Fig:Improvement}
\end{figure*}
Finally in the third column of Fig.~\ref{fig:eff} considers the effect on lepton (photon) triggers on the process with additional $W/Z$ (photon) radiation. 
The two models are almost identical, showing that these triggers are picking up the extra electroweak radiation. This assures us that the dark pion lifetimes and parameters of the confined hidden sector have little bearing on the EW focus of the lepton triggers. The slightly higher efficiency for lepton triggers ($e, \mu$) in Model \textbf{A} because of the dark pion long lifetimes leaving less contaminants within the lepton isolation cone, i.e Eq.~(\ref{eqn:Icone}). In terms of the dependence on $M_{Z'}$, the electroweak radiation is roughly constant, while the photons tend to be harder at larger $M_{Z'}$ so the efficiency increases. 

Although a useful metric, the efficiency does not consider the overall probability of the event occurring because it does not take into account that extra radiation reduces the cross section. Therefore when looking at the total rate, we consider the cross section of the process times its respective efficiency. To get a proper sense of the rate independent of some of the unknown particle physics such as the couplings of the $Z'$, we take the ratio of the cross section times efficiency with respect to the leading order (LO) result with no additional radiation, and show the results in Fig.~\ref{Fig:Improvement}. In the left column we show the improvement achieved for the MET trigger in Model \textbf{A}. We see that QCD radiation can lead to significant improvements at low mass, and even at high mass simulating extra radiation increases the overall rate by $\mathcal{O}(100)$\%. We also see that simulating two additional jets gives significant improvement relative to only a single jet at low mass. Electroweak radiation only gives a modest improvement in the event rate, roughly 10\% at low mass and even more modest at high mass. This is because the rate suppression due to $\alpha$ is significantly stronger than that from QCD that goes like $\alpha_s$. 

In the right column of Fig.~\ref{Fig:Improvement} we show the improvement for the $H_{T}$ trigger in Model \textbf{B}. At low mass, none of the $400$ $000$ events we simulate at leading order pass the trigger, so considering radiation opens a new parameter regime for discovery. Even at intermediate masses, $M_{Z'} \sim 500$ GeV, additional radiation gives orders of magnitude improvement in rate. As with Model \textbf{A}, two additional jets gives the greatest improvement, but all processes considered can be significant.

\section{Machine learning triggers}
\label{sec:ML}

The current triggers of Section~\ref{sec:current}, although well understood, are still limited in their capacity to find new physics. They have forced us to consider subsets of model parameter space that best produce specific signals that current triggers discriminate best on and create blind spots in other regions of parameter space. The CMS and ATLAS experiments have been investigating more novel methods of ML analyses at the trigger level~\cite{Alimena_2020}. These non-linear algorithms are not limited to primary detector features, such as hard jets and charged particle tracks, as they can be trained (unsupervised) to converge on non-intuitive abstract features within an event. These algorithms can be trained on a array of new physics simulations, opening up the door for unique correlations between new physics models that could prove generic for a dedicated trigger. Additionally, the computational resources necessary in training and testing the ML method does not impact the resources in-situ as the training/testing stage is done prior to implementation on the trigger stream.

Emerging jets produce interesting signatures and are complimentary to other models consisting of hidden sector portals with dark showers. Examples include semi-visible jets~\cite{Cohen_2015, Cohen_2017} and SUEPS~\cite{Knapen_2017}, which exhibit extreme cases of a similar baseline theory to that of emerging jets. In this section, we employ the use of the ATLAS inner trackers in training ML methods for emerging jet signals for the purpose of triggering. Like the complimentary models mentioned above, emerging jets produce a wealth of uncharacteristic lifetimes as compared to the SM. Here we show that recruiting the inner tracker allows an ML algorithm to converge on discriminating features that exploit this gap in lifetimes between the new physics and the SM. Whether ML methods converge on generic features of new physics or those more specific to the training model, understanding what is the physics behind these ML features has been a long standing question. Answering this question requires solving an inverse problem of the ML output, and proves more and more difficult as ML methods such as neural networks and boosted decision trees become more complex, although there is some recent progress~\cite{Faucett:2020vbu}. Fortunately we can largely disregard this problem since the purpose of the trigger is to get as many interesting events onto record as possible. Anything written onto record can then be properly analyzed offline. So we take the pragmatic approach to ML and focus our attention on producing the highest trigger efficiency independent of what it is ``seeing.''

As in Section~\ref{sec:current}, the trigger systems at both L1 and HLT are limited by their allocated computational resources. These triggering operations must be fast enough to reduce the input data stream to $\sim1$ kHz. This can be challenging when data from all areas of the detector package are simultaneously used within the triggering systems in some form or another. Fitting non-linear algorithms at the trigger level, such as ML methods can be taxing on the available resources. In this case, we look at low level variables such as hits on the tracking detectors and simple jet reconstructions from L1. The lack of fully constructed particle tracks and momentum measurements allows for fully trained algorithms to operate quickly on the incoming data streams.

\begin{figure}
\centering 
\includegraphics[width=.48\textwidth ]{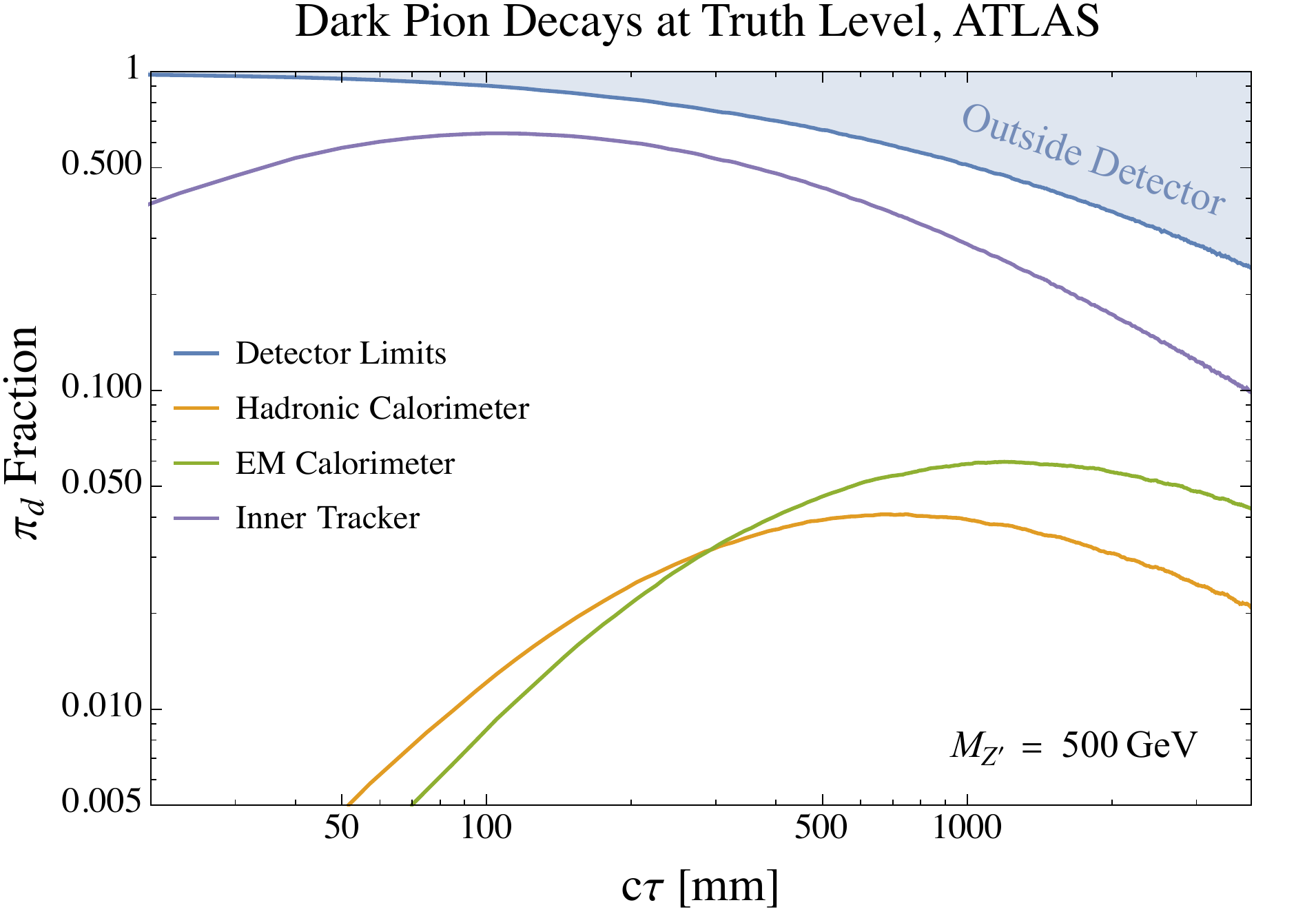}
\caption{The fraction of dark pions that decay into visible sector particles within the primary ATLAS detectors. The green line represents the decay fraction within the electromagnetic calorimeter, the orange for the hadronic calorimeter, and the blue for the inner trackers. The blue line is the fiducial boundary of the muon spectrometer which we take to be the boundary of the ATLAS detector. We do not show that fraction that decay before the first tracker layer, and this fraction is significant at short lifetime. These fractions are computed from simulation with $M_{Z'} = 500$ GeV.}
\label{fig:lifetime}
\end{figure}

For concreteness, we will analyze the ATLAS tracker geometry in this section, but we expect similar qualitative conclusions for CMS. 
In Fig.~\ref{fig:lifetime} we show the fraction of the dark pions that decay in a given detector subsystem as a function of lifetime, and we see that for all lifetimes, the largest fraction is in the inner tracker. Therefore, we focus only on that system in this work, though we note that there could be interesting improvements by including the calorimeters and muon system.

The ATLAS inner tracker comprises of, in increasing order of distance from the beam pipe: the Inner Bi-layer (IBL), the Pixel Detectors, the silicon detectors, and the Transition Radiation Tracker (TRT). Table~\ref{tab:tracker}
shows the number of layers and radial distances from the interaction point as well as whether the tracker has additional endcaps. The primary purpose of the inner trackers is track reconstruction and particle identification. Tracks are reconstructed at the HLT level where calorimeter information has been seeded from L1. Although useful, track reconstruction is a very computationally expensive operation as all possible track trajectories are back propagated towards the interaction point. We propose using machine learning methods trained on the hit patterns of the inner tracker's layers while bypassing full track reconstruction. 

As noted in Section~\ref{sec:event}, in our simulations we take the dark hadronization to be dominated by dark pions. The proper lifetime of the dark pions depend strongly on the $Z'$ mass as shown in Eq.~(\ref{eq:lifetime}), with lower mediator masses corresponding to shorter lifetimes. We assume a common lifetime for the dark pions, but for a study of hierarchical lifetimes see~\cite{Renner:2018fhh}, and we take the lifetime to be a free parameter that we vary.

As the emerging jets traverse the inner tracker, the invisible dark pions decay into visible quarks which hadronize into SM jets with high particle multiplicity creating a complex bundle of displaced tracks as they decay throughout the detector volume. 
Ideally, without pileup and other secondary detector effects, the number of hits registered on the tracker layers should increase with radial distance. 

\begin{table}
\centering
\begin{tabular}{cccc}
\hline
Tracker & Layer &  Radius (mm) & Geometry \\ \hline 
IBL &      & 33.25  & Barrel \\ \hline
         & First  &  50.5 &\\
Pixel  & Second  & 88.5 & Barrel + Endcap  \\ 
        & Third  & 122.5 & \\  \hline
        & First  & 299 &  \\ 
SCT  & Second  & 371 &  Barrel + Endcap \\   
        & Third & 443 &\\ 
  & Fourth &  514 &\\ \hline
TRT &  Start & 554 & Barrel + Endcap\\ \hline
\end{tabular}
\caption{ATLAS inner tracker specifications taken from ~\citep{Airapetian:391176,Capeans:1291633}. The barrel layers of each tracker section are shown with their radial distance from the beam pipe/interaction point. Since the TRT is a more complex tracking package, we only consider the hits on the initial layer of the TRT. Endcaps also accompany most of the trackers but aren't considered in this analysis.  }
\label{tab:tracker}
\end{table}

Signal and background events are simulated and explained in Section~\ref{sec:event}. Hadronized events are passed to a highly simplified detector simulation used for~\cite{ Knapen_2017} that attempts to model the ATLAS inner tracker. The simulation accounts for updated particle trajectories from the bending of the ATLAS toroidal magnetic field, $B \approx 2$T, as well as energy loss from interactions with the material (assuming a thin layer approximation). The simulation does not account for the production of secondary particles from interactions with the layer materials. These secondaries are a possible source of error as they could fake a displaced vertex around each tracker. We take combinations of the concentric tracker volumes utilizing the IBL$_{i}$, Pixel$_{i}$, SCT$_{i}$ and TRT$_{i}$ detectors where the subscript denotes the $i^{th}$ layer. Results are represented as hit patterns in the form of heat maps in the ($\phi$, $ \eta$) plane for each layer of the trackers. 

A simple strategy that ultimately does not work is to train a classifier using only the total number of hits recorded on each layer. This strategy reduces the 2D map in ($\phi, \eta$) space to a single variable $N_{h,i}$ representing the number of hits on the $i^{th}$ layer. ROOT's TMVA machine learning toolkit~\citep{Hocker:2007ht} is used to train and test the ML algorithm, in this case a support vector machine, with both signal and $b\bar{b}$ background events. The primary reason this strategy does not work is because of pileup. Each LHC event contains pileup from multiple proton collisions in each event. While there is only one hard collision responsible for the emerging jet production, the remaining collisions produce a large number of soft particles distributed approximately isotropically through the detector. This will typically wash out large differences in hits on consecutive layers $ N_{h,i} - N_{h, i + 1} \approx 0$. We simulate pileup by adding an average of 50 minimum bias collisions to both signal and $b \bar{b}$ background events. Since pileup is relatively soft, the toroid magnet will deflect charged particles with a radius of curvature inversely proportional to the particles momentum $R_{\textrm{c}} \propto p^{-1} $. Pileup, in the form of minimum bias events, has characteristically low momentum spectrum, so layers at larger radial distances are less sensitive to pileup effects.

To minimize the influence of pileup, we refine our strategy by adding a simplified jet reconstruction algorithm such that instead of counting hits on the entire layer, we instead count tracker hits in the geometric vicinity of hard jets. In detail, energy deposition in the calorimeter is used to reconstruct jets in terms of topo clusters at L1 and at HLT. With jet information at the trigger levels we use the truth level jet vectors $\hat{v}$ as well as the jet cone acceptance $\mathcal{R} = 0.5$ to define $N_{cor}^i$, 
\begin{equation}\label{eq:ncor}
N_{cor}^i \equiv \sum_{j} N_{h,j}^i\left( \sqrt{(\Delta\eta)_{j}^2 +(\Delta \phi)_{j}^2} \leq \mathcal{R} \right).
\end{equation}
Where the sum runs through all grid points $j$ of the layer $i$. $(\Delta\eta)_{j} = \eta_{j} - \eta_{\hat{v}}$, where $\eta_{j}$ are pseudorapidities at grid point j and $ \eta_{\hat{v}}$ is the pseudorapidity of leading jet direction $\hat{v}$. Similarly, $(\Delta\phi)_{j} = \phi_{j} - \phi_{\hat{v}}$. The radius of each layer $R_{i}$ values can be found in Table~\ref{tab:tracker}. This approach allows the classifier to become sensitive to a conical subset of hit patterns in the direction of an L1 topocluster jets. This substantially reduces the $4 \pi $ reach of pileup.

\begin{figure*}
\centering
\begin{minipage}[c]{\textwidth}
\includegraphics[width=.49\textwidth ]{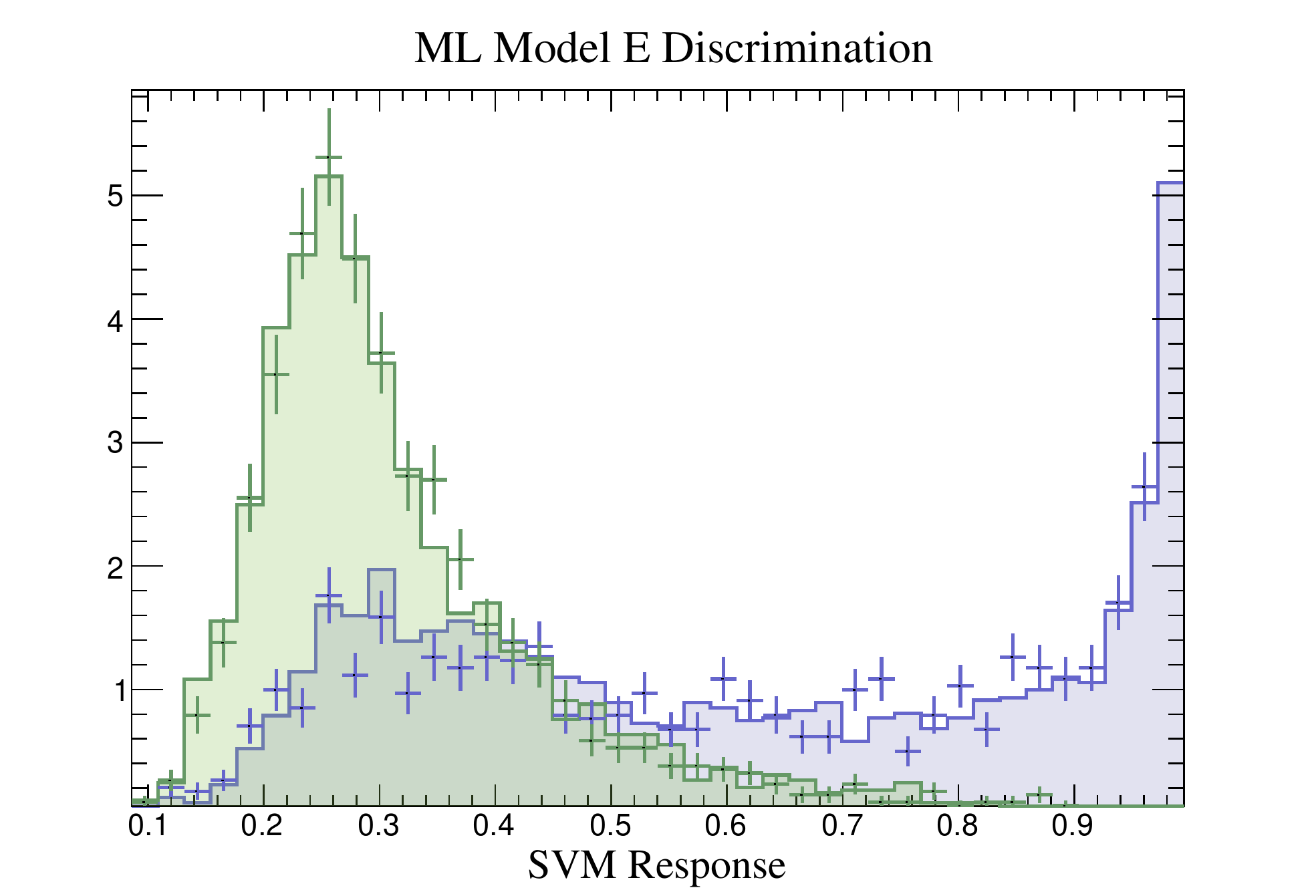}
\hfill
\includegraphics[width=.49\textwidth ]{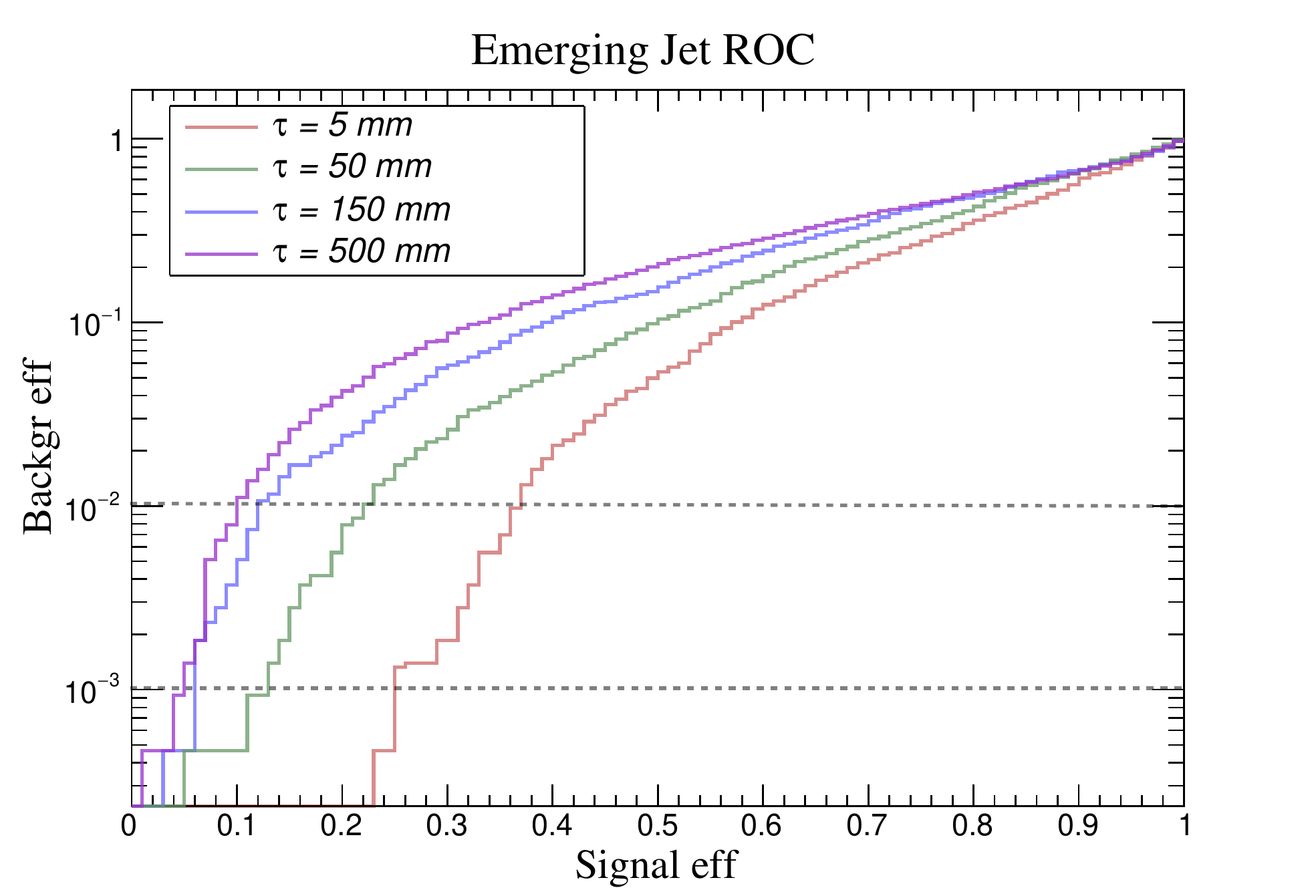} 
\end{minipage}
\hfill
\caption{On the left, Model E (blue) discrimination from $b\bar{b}$ background (green) using a support vector machine. The SVM was trained using four layers IBL, Pixel$_{2}$, SCT$_{2}$, TRT$_{1}$. An average pileup of $\mu = 50$ was added to both signal and background. The flat bars (points) correspond to the training (test) set. On the right, Receiver Operation Characteristic ROC for Models \textbf{A}, \textbf{C},\textbf{D}, and \textbf{E} using the TMVA support vector machine. From Eq.~(\ref{eq:ncor}) the SVM was trained and tested using four layers: IBL, Pixel$_{2}$, SCT$_{2}$, TRT$_{1}$. At a given background efficiency the expected signal efficiencies increase as the dark pion lifetimes lower. The required background rejection is estimated to lie between the horizontal dotted lines.}
\label{fig:ml}

\end{figure*}

For this analysis, we fix $M_{Z'} = 500$ GeV. Models \textbf{A}, \textbf{C}, \textbf{D} and \textbf{E} (described in Table~\ref{tab:benchmarks}) are used as signal benchmarks. They vary in only the lifetimes while keeping all other dark sector parameters equal. We also included Model \textbf{F}, which has the lifetime of Model \textbf{D} but varies in the dark sector parameters. The TMVA support vector machine was trained and tested using Eq.~(\ref{eq:ncor}) as the input variables. Testing and training sets were created and randomly separated from the simulation results. The left panel of Fig.~\ref{fig:ml} shows the resulting SVM signal and background discrimination. We see that there is excellent separation between signal background, and that the trained and tested distributions look very similar. 
Only four layers of the trackers were used, IBL, Pixel$_{2}$, SCT$_{2}$, TRT$_{1}$, sampled from each of the four inner tracker packages. This arrangement of layers allows the ML to train on snapshots of the emerging jets evolution at substantially separated intervals, but using only four layers reduces the required size of the training sample. A proper study could simulate a much larger and realistic sample size while incorporating additional layers. 

Choosing a value for the SVM response dictates the Type I (false positive) and Type II (false negative) errors which are related to the power of the signal/background classification. Assuming that all events all background saturated, meaning very little new physics events occur, implemented triggers must reach background rates that do not exceed the allocated bandwidth. Unlike in Section~\ref{sec:current} where we investigated pre-existing triggers, the background rates of our proposed novel triggers are unknown and must be estimated. The high level trigger rate usually allocated for new triggers is of $R \sim 1$ Hz~\citep{2013LLPTrigger}. The required background rejection is given by,
\begin{equation}\label{eq:rate}
\epsilon_{\textrm{bkg}} = \frac{R}{\sigma_{\textrm{bkg}} \mathcal{L}} \sim 10^{-3} - 10^{-2}
\end{equation}
for a peak instantaneous luminosity of $\mathcal{L} = 21\times 10^{33} \textrm{ cm}^{-2}\textrm{s}^{-1}$. The background rates were estimated assuming high $p_T$ $b\bar{b}$ production, which primarily mimic the signal events. Additional backgrounds were considered, such as inclusive hard QCD backgrounds, generated through \verb!Pythia8!. Since these additional background sources had substantially smaller efficiencies than pure heavy flavour $b\bar{b}$, we take them to be negligible to the total background rates. Inclusive background cross sections were taken from the \verb!Pythia8! hard event generation as explained in Section~\ref{sec:event}, but these background cross sections are leading order and thus only order of magnitude estimates.

The Type I and Type II errors of Fig.~\ref{fig:ml} are closely related to the signal and background efficiencies. A more robust visualization of this relationship is seen when plotting the Receiver Operator Characteristic curve (ROC). The ROC curve is calculated by plotting the background vs.~signal efficiency for each SVM response value as shown in the right panel of Fig.~\ref{fig:ml}. The value of the signal efficiency $\epsilon$ can be read off for each model at the highest allowable ATLAS trigger background efficiency. In Table~\ref{tab:BkgRej} we see the range of $\epsilon$ for various lifetimes ranging from $5$ mm to $500$ mm. Shorter lifetimes seem to out perform longer lifetimes. In Model \textbf{E}  ($c\tau_d = 5$ mm) there are distinct hit patterns all within the IBL and the beginning of the TRT. These dark pions are boosted such that their respective emerging jets must go through majority of its evolution between these inner tracker slices. Whereas for the higher lifetime models, their evolution from invisible to visible is either skewed towards the later layers or even clipped beyond the tracker limits. Efficiencies of $\mathcal{O}(0.1 - 0.3)$ can be found for a comparable resonance mass of $M_{Z'} = 500$ GeV for the simple $s$-channel process with no additional hard ISR. Comparing these to the signal efficiencies found in Section~\ref{sec:current} (see Fig.~\ref{fig:eff}) we see that for similar topologies, training on detector hits can be advantageous as current triggers have less reach in the low mass regime. 

\vspace{1em}
\begin{table}[b]
\centering
\begin{tabular}{ccc}
\hline
 $c\tau_{d}$ & $\epsilon$ ( Bkg rej $10^{-2})$ & $\epsilon$ (Bkg rej $10^{-3})$ \\ \hline 
5  mm & 0.370 & 0.250 \\
50  mm & 0.230  & 0.125 \\
150 mm & 0.122  & 0.060 \\
500 mm & 0.100  & 0.050  \\ \hline
\end{tabular}
\caption{Signal efficiencies for expected allowable background rates for new ATLAS triggers for $M_{Z'} = 500$ GeV. The lifetimes represent Models \textbf{A}, \textbf{C}, \textbf{D}, and  \textbf{E}. Each value in the last two columns are extracted from Fig.~\ref{fig:ml} for a bkg efficiency calculated using Eq.~(\ref{eq:rate}) assuming the events are fully saturated by background. }
\label{tab:BkgRej}
\end{table}

\begin{figure}
 
\makebox[\textwidth][l]{\includegraphics[width=.48\textwidth]{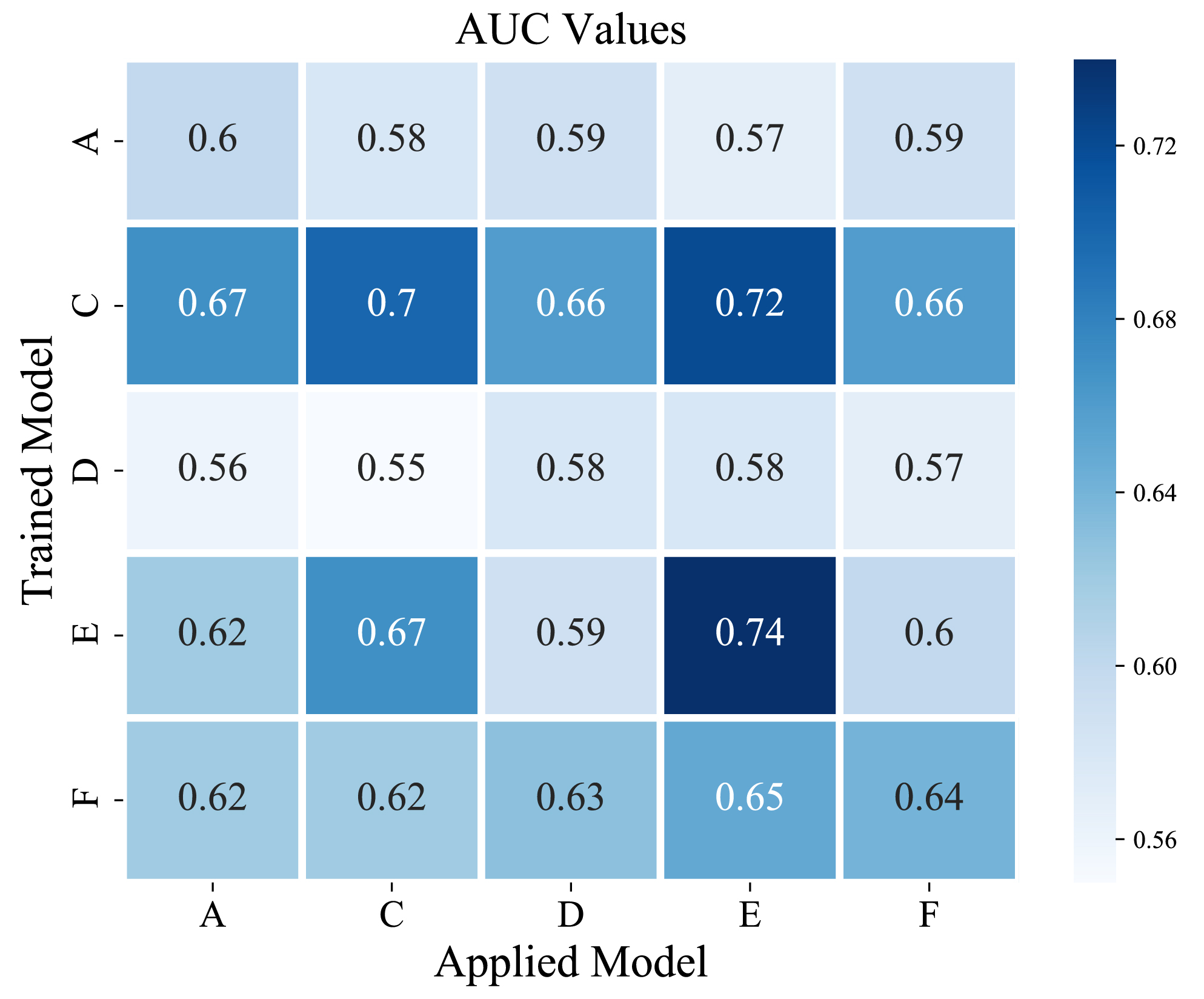} }
\caption{Area Under the Curve (AUC) of the ROC curves. The SVM was trained on specific models (vertical axis) and tested/applied on a unknown data set (horizontal axis) of the same/alternative model. The diagonals are the AUC values corresponding for training and testing on the same models i.e. the ROC curves of Fig.~\ref{fig:ml}. Trained models show little separation in classification power when applied to a range of lifetimes and hidden sector parameters.}
\label{fig:AUC}
\end{figure}

So far we have discussed training and testing on the same model parameters. Implementing an ML trigger would require training on some expected signal model prior to integration on the trigger stream. As mentioned earlier, emerging jets and other models of dark showers have a large available lifetime parameter space. Training an ML trigger on a single model parameter point could bias the trigger towards classifying only a small portion of this parameter space. The overall classification power is related to the area under the ROC curve (AUC). As a test of the universality of this method, we take an unknown sample set from each of the five models and apply each of the trained SVMs on them as was done in~\cite{Cesarotti:2020hwb}. The results are seen in where the diagonal corresponds to the AUC of Fig.~\ref{fig:AUC}. Each row is of an SVM trained on a single lifetime and then applied to an unknown signal set of differing or same lifetime (columns). The deviations from each unknown lifetime set is of order a few percent. This reinforces the insensitivity of the trigger to the parameters it was trained on. Each rows' average AUC value does not have a substantial change. Model \textbf{F} was included to see how sensitive this analysis is on the parameters of the hidden sector, such as the hadronization scale $\Lambda_{d}$, and the dark composite masses ($\pi_{d}, \rho_{d}$). The first thing to notice is that when trained on Model \textbf{F} there is almost no change in the AUC when applied to the range of lifetime models, much like Model \textbf{D} which shares the same lifetime. Secondly, when the various trained models are applied to both Model \textbf{F} and Model \textbf{D} the AUC values are almost identical. This similarity gives us more confidence in the universality of the trigger, such that it is mostly sensitive to pion lifetimes while being largely insensitive to other deviations in the hidden sector.  

\section{Summary and conclusions}
\label{sec:conclusion}

In this work we explore ways to trigger on new physics models with a high multiplicity of displaced vertex of the type explored in~\cite{Schwaller:2015gea}. If the mediator is uncharged under the SM gauge groups, then it can be relatively light and associated SM radiation is not guaranteed, so triggering on these events at the LHC is not straightforward. If events are not triggered on, they are forever lost, so maximizing trigger efficiency is a necessary step for maximizing discovery potential. 

We have first explored how both efficiency and event rate can be increased using current triggers with the addition of initial state radiation. Our main results are given in Figs.~\ref{fig:eff} and~\ref{Fig:Improvement}. Both QCD and electroweak radiation increase the total energy of the event and thus increases the efficiency of $H_{T}$ triggers. The effects are the largest for relatively light mediators, but they can also be relevant for heavy mediators. Other triggers, such as those searching for missing energy, leptons, and photons can also have higher efficiency with certain radiation. Processes with radiation have lower cross sections than the leading order process, but we have shown that even taking this into account, there are significant increases in event rate, especially at low mass. This ISR process is guaranteed by gauge invariance to exist, and we encourage experimentalists to simulate these processes in future studies.

We have also explored possible new triggers using modern machine learning methods that use simple counts of hits in the tracker as input variables. The new physics models considered here leave an increasing number of hits on each tracker layer as the dark mesons decay in flight to visible SM states. Counting hits in the tracker is significantly faster than performing track reconstruction, making it an ideal technique for a trigger. The use of ML techniques allows for a more sophisticated separation of signal and background based on these hit counts, and the effectiveness of our proposed method is shown in Fig.~\ref{fig:ml}. Using ML for triggering can significantly reduce one of the primary problems of ML techniques in particle physics that it is difficult to determine which features the algorithms are training on, and thus difficult to estimate systematic errors. For a trigger, one wants to maximize the events recorded, and then a full study of systematics can be done at the analysis stage. 

Finally, we have explored the sensitivity of our ML techniques across different model parameters as summarized in Fig.~\ref{fig:AUC}. Of course, the underlying model parameters of the new physics are unknown, and the ideal trigger would be sensitive to as much of the parameter space as possible and also to new physics models not captured by the simulation framework used in this work. We see that varying the particle physics parameters of the dark sector does significantly affect the efficiency, so a realistic trigger can be trained on one model parameter and still be sensitive to a broad class of new physics models. 

~\\
\noindent
{\bf Acknowledgments:}~We thank Paul Archer-Smith, Kevin Earl, Dag Gillberg, Simon Knapen, Siddharth Mishra-Sharma, Michael Ramsey-Musolf, Pedro Schwaller, Jesse Thaler, and Andreas Weiler for helpful discussions. This work is supported in part by the Natural Sciences and Engineering Research Council of Canada (NSERC). DL acknowledges support from the NSERC Postgraduate Scholarships Doctoral Program (PGS D). DS is grateful for the hospitality from the Kavli Institute for Theoretical Physics and the Galileo Galilei Institute for Theoretical Physics where part of this work was done.


\bibliographystyle{apsrev}
\bibliography{ref}

\begin{thebibliography}{69}
\expandafter\ifx\csname natexlab\endcsname\relax\def\natexlab#1{#1}\fi
\expandafter\ifx\csname bibnamefont\endcsname\relax
  \def\bibnamefont#1{#1}\fi
\expandafter\ifx\csname bibfnamefont\endcsname\relax
  \def\bibfnamefont#1{#1}\fi
\expandafter\ifx\csname citenamefont\endcsname\relax
  \def\citenamefont#1{#1}\fi
\expandafter\ifx\csname url\endcsname\relax
  \def\url#1{\texttt{#1}}\fi
\expandafter\ifx\csname urlprefix\endcsname\relax\def\urlprefix{URL }\fi
\providecommand{\bibinfo}[2]{#2}
\providecommand{\eprint}[2][]{\url{#2}}

\bibitem[{\citenamefont{Schwaller et~al.}(2015)\citenamefont{Schwaller,
  Stolarski, and Weiler}}]{Schwaller:2015gea}
\bibinfo{author}{\bibfnamefont{P.}~\bibnamefont{Schwaller}},
  \bibinfo{author}{\bibfnamefont{D.}~\bibnamefont{Stolarski}},
  \bibnamefont{and} \bibinfo{author}{\bibfnamefont{A.}~\bibnamefont{Weiler}},
  \bibinfo{journal}{JHEP} \textbf{\bibinfo{volume}{05}}, \bibinfo{pages}{059}
  (\bibinfo{year}{2015}), \eprint{1502.05409}.

\bibitem[{\citenamefont{Cohen et~al.}(2015)\citenamefont{Cohen, Lisanti, and
  Lou}}]{Cohen_2015}
\bibinfo{author}{\bibfnamefont{T.}~\bibnamefont{Cohen}},
  \bibinfo{author}{\bibfnamefont{M.}~\bibnamefont{Lisanti}}, \bibnamefont{and}
  \bibinfo{author}{\bibfnamefont{H.~K.} \bibnamefont{Lou}},
  \bibinfo{journal}{Phys. Rev. Lett.} \textbf{\bibinfo{volume}{115}},
  \bibinfo{pages}{171804} (\bibinfo{year}{2015}), \eprint{1503.00009}.

\bibitem[{\citenamefont{Csaki et~al.}(2015)\citenamefont{Csaki, Kuflik,
  Lombardo, and Slone}}]{Csaki:2015fba}
\bibinfo{author}{\bibfnamefont{C.}~\bibnamefont{Csaki}},
  \bibinfo{author}{\bibfnamefont{E.}~\bibnamefont{Kuflik}},
  \bibinfo{author}{\bibfnamefont{S.}~\bibnamefont{Lombardo}}, \bibnamefont{and}
  \bibinfo{author}{\bibfnamefont{O.}~\bibnamefont{Slone}},
  \bibinfo{journal}{Phys. Rev. D} \textbf{\bibinfo{volume}{92}},
  \bibinfo{pages}{073008} (\bibinfo{year}{2015}), \eprint{1508.01522}.

\bibitem[{\citenamefont{Cohen et~al.}(2017)\citenamefont{Cohen, Lisanti, Lou,
  and Mishra-Sharma}}]{Cohen_2017}
\bibinfo{author}{\bibfnamefont{T.}~\bibnamefont{Cohen}},
  \bibinfo{author}{\bibfnamefont{M.}~\bibnamefont{Lisanti}},
  \bibinfo{author}{\bibfnamefont{H.~K.} \bibnamefont{Lou}}, \bibnamefont{and}
  \bibinfo{author}{\bibfnamefont{S.}~\bibnamefont{Mishra-Sharma}},
  \bibinfo{journal}{JHEP} \textbf{\bibinfo{volume}{11}}, \bibinfo{pages}{196}
  (\bibinfo{year}{2017}), \eprint{1707.05326}.

\bibitem[{\citenamefont{Knapen et~al.}(2017)\citenamefont{Knapen, Pagan~Griso,
  Papucci, and Robinson}}]{Knapen_2017}
\bibinfo{author}{\bibfnamefont{S.}~\bibnamefont{Knapen}},
  \bibinfo{author}{\bibfnamefont{S.}~\bibnamefont{Pagan~Griso}},
  \bibinfo{author}{\bibfnamefont{M.}~\bibnamefont{Papucci}}, \bibnamefont{and}
  \bibinfo{author}{\bibfnamefont{D.~J.} \bibnamefont{Robinson}},
  \bibinfo{journal}{JHEP} \textbf{\bibinfo{volume}{08}}, \bibinfo{pages}{076}
  (\bibinfo{year}{2017}), \eprint{1612.00850}.

\bibitem[{\citenamefont{Pierce et~al.}(2018)\citenamefont{Pierce, Shakya, Tsai,
  and Zhao}}]{Pierce:2017taw}
\bibinfo{author}{\bibfnamefont{A.}~\bibnamefont{Pierce}},
  \bibinfo{author}{\bibfnamefont{B.}~\bibnamefont{Shakya}},
  \bibinfo{author}{\bibfnamefont{Y.}~\bibnamefont{Tsai}}, \bibnamefont{and}
  \bibinfo{author}{\bibfnamefont{Y.}~\bibnamefont{Zhao}},
  \bibinfo{journal}{Phys. Rev. D} \textbf{\bibinfo{volume}{97}},
  \bibinfo{pages}{095033} (\bibinfo{year}{2018}), \eprint{1708.05389}.

\bibitem[{\citenamefont{Burdman and Lichtenstein}(2018)}]{Lichtenstein:2018kno}
\bibinfo{author}{\bibfnamefont{G.}~\bibnamefont{Burdman}} \bibnamefont{and}
  \bibinfo{author}{\bibfnamefont{G.}~\bibnamefont{Lichtenstein}},
  \bibinfo{journal}{JHEP} \textbf{\bibinfo{volume}{08}}, \bibinfo{pages}{146}
  (\bibinfo{year}{2018}), \eprint{1807.03801}.

\bibitem[{\citenamefont{Cheng et~al.}(2019)\citenamefont{Cheng, Li, Salvioni,
  and Verhaaren}}]{Cheng:2019yai}
\bibinfo{author}{\bibfnamefont{H.-C.} \bibnamefont{Cheng}},
  \bibinfo{author}{\bibfnamefont{L.}~\bibnamefont{Li}},
  \bibinfo{author}{\bibfnamefont{E.}~\bibnamefont{Salvioni}}, \bibnamefont{and}
  \bibinfo{author}{\bibfnamefont{C.~B.} \bibnamefont{Verhaaren}},
  \bibinfo{journal}{JHEP} \textbf{\bibinfo{volume}{11}}, \bibinfo{pages}{031}
  (\bibinfo{year}{2019}), \eprint{1906.02198}.

\bibitem[{\citenamefont{Alimena et~al.}(2020{\natexlab{a}})}]{Alimena_2020_ds}
\bibinfo{author}{\bibfnamefont{J.}~\bibnamefont{Alimena}} \bibnamefont{et~al.},
  \bibinfo{journal}{J. Phys. G} \textbf{\bibinfo{volume}{47}},
  \bibinfo{pages}{090501} (\bibinfo{year}{2020}{\natexlab{a}}),
  \eprint{1903.04497}.

\bibitem[{\citenamefont{Strassler and Zurek}(2007)}]{Strassler:2006im}
\bibinfo{author}{\bibfnamefont{M.~J.} \bibnamefont{Strassler}}
  \bibnamefont{and} \bibinfo{author}{\bibfnamefont{K.~M.} \bibnamefont{Zurek}},
  \bibinfo{journal}{Phys. Lett.} \textbf{\bibinfo{volume}{B651}},
  \bibinfo{pages}{374} (\bibinfo{year}{2007}), \eprint{hep-ph/0604261}.

\bibitem[{\citenamefont{Strassler and Zurek}(2008)}]{Strassler:2006ri}
\bibinfo{author}{\bibfnamefont{M.~J.} \bibnamefont{Strassler}}
  \bibnamefont{and} \bibinfo{author}{\bibfnamefont{K.~M.} \bibnamefont{Zurek}},
  \bibinfo{journal}{Phys. Lett. B} \textbf{\bibinfo{volume}{661}},
  \bibinfo{pages}{263} (\bibinfo{year}{2008}), \eprint{hep-ph/0605193}.

\bibitem[{\citenamefont{Han et~al.}(2008)\citenamefont{Han, Si, Zurek, and
  Strassler}}]{Han:2007ae}
\bibinfo{author}{\bibfnamefont{T.}~\bibnamefont{Han}},
  \bibinfo{author}{\bibfnamefont{Z.}~\bibnamefont{Si}},
  \bibinfo{author}{\bibfnamefont{K.~M.} \bibnamefont{Zurek}}, \bibnamefont{and}
  \bibinfo{author}{\bibfnamefont{M.~J.} \bibnamefont{Strassler}},
  \bibinfo{journal}{JHEP} \textbf{\bibinfo{volume}{07}}, \bibinfo{pages}{008}
  (\bibinfo{year}{2008}), \eprint{0712.2041}.

\bibitem[{\citenamefont{Bai and Schwaller}(2014)}]{Bai_2014}
\bibinfo{author}{\bibfnamefont{Y.}~\bibnamefont{Bai}} \bibnamefont{and}
  \bibinfo{author}{\bibfnamefont{P.}~\bibnamefont{Schwaller}},
  \bibinfo{journal}{Phys. Rev. D} \textbf{\bibinfo{volume}{89}},
  \bibinfo{pages}{063522} (\bibinfo{year}{2014}), \eprint{1306.4676}.

\bibitem[{\citenamefont{Lonsdale et~al.}(2017)\citenamefont{Lonsdale, Schroor,
  and Volkas}}]{Lonsdale:2017mzg}
\bibinfo{author}{\bibfnamefont{S.~J.} \bibnamefont{Lonsdale}},
  \bibinfo{author}{\bibfnamefont{M.}~\bibnamefont{Schroor}}, \bibnamefont{and}
  \bibinfo{author}{\bibfnamefont{R.~R.} \bibnamefont{Volkas}},
  \bibinfo{journal}{Phys. Rev. D} \textbf{\bibinfo{volume}{96}},
  \bibinfo{pages}{055027} (\bibinfo{year}{2017}), \eprint{1704.05213}.

\bibitem[{\citenamefont{Lonsdale and Volkas}(2018)}]{Lonsdale:2018xwd}
\bibinfo{author}{\bibfnamefont{S.~J.} \bibnamefont{Lonsdale}} \bibnamefont{and}
  \bibinfo{author}{\bibfnamefont{R.~R.} \bibnamefont{Volkas}},
  \bibinfo{journal}{Phys. Rev. D} \textbf{\bibinfo{volume}{97}},
  \bibinfo{pages}{103510} (\bibinfo{year}{2018}), \eprint{1801.05561}.

\bibitem[{\citenamefont{Craig et~al.}(2015)\citenamefont{Craig, Katz,
  Strassler, and Sundrum}}]{Craig:2015pha}
\bibinfo{author}{\bibfnamefont{N.}~\bibnamefont{Craig}},
  \bibinfo{author}{\bibfnamefont{A.}~\bibnamefont{Katz}},
  \bibinfo{author}{\bibfnamefont{M.}~\bibnamefont{Strassler}},
  \bibnamefont{and} \bibinfo{author}{\bibfnamefont{R.}~\bibnamefont{Sundrum}},
  \bibinfo{journal}{JHEP} \textbf{\bibinfo{volume}{07}}, \bibinfo{pages}{105}
  (\bibinfo{year}{2015}), \eprint{1501.05310}.

\bibitem[{\citenamefont{Curtin and Verhaaren}(2015)}]{Curtin:2015fna}
\bibinfo{author}{\bibfnamefont{D.}~\bibnamefont{Curtin}} \bibnamefont{and}
  \bibinfo{author}{\bibfnamefont{C.~B.} \bibnamefont{Verhaaren}},
  \bibinfo{journal}{JHEP} \textbf{\bibinfo{volume}{12}}, \bibinfo{pages}{072}
  (\bibinfo{year}{2015}), \eprint{1506.06141}.

\bibitem[{\citenamefont{Cheng et~al.}(2017)\citenamefont{Cheng, Salvioni, and
  Tsai}}]{Cheng:2016uqk}
\bibinfo{author}{\bibfnamefont{H.-C.} \bibnamefont{Cheng}},
  \bibinfo{author}{\bibfnamefont{E.}~\bibnamefont{Salvioni}}, \bibnamefont{and}
  \bibinfo{author}{\bibfnamefont{Y.}~\bibnamefont{Tsai}},
  \bibinfo{journal}{Phys. Rev. D} \textbf{\bibinfo{volume}{95}},
  \bibinfo{pages}{115035} (\bibinfo{year}{2017}), \eprint{1612.03176}.

\bibitem[{\citenamefont{Kilic et~al.}(2019)\citenamefont{Kilic, Najjari, and
  Verhaaren}}]{Kilic:2018sew}
\bibinfo{author}{\bibfnamefont{C.}~\bibnamefont{Kilic}},
  \bibinfo{author}{\bibfnamefont{S.}~\bibnamefont{Najjari}}, \bibnamefont{and}
  \bibinfo{author}{\bibfnamefont{C.~B.} \bibnamefont{Verhaaren}},
  \bibinfo{journal}{Phys. Rev. D} \textbf{\bibinfo{volume}{99}},
  \bibinfo{pages}{075029} (\bibinfo{year}{2019}), \eprint{1812.08173}.

\bibitem[{\citenamefont{Park and Zhang}(2019)}]{Park:2017rfb}
\bibinfo{author}{\bibfnamefont{M.}~\bibnamefont{Park}} \bibnamefont{and}
  \bibinfo{author}{\bibfnamefont{M.}~\bibnamefont{Zhang}},
  \bibinfo{journal}{Phys. Rev. D} \textbf{\bibinfo{volume}{100}},
  \bibinfo{pages}{115009} (\bibinfo{year}{2019}), \eprint{1712.09279}.

\bibitem[{\citenamefont{Cohen et~al.}(2020)\citenamefont{Cohen, Doss, and
  Freytsis}}]{Cohen:2020afv}
\bibinfo{author}{\bibfnamefont{T.}~\bibnamefont{Cohen}},
  \bibinfo{author}{\bibfnamefont{J.}~\bibnamefont{Doss}}, \bibnamefont{and}
  \bibinfo{author}{\bibfnamefont{M.}~\bibnamefont{Freytsis}},
  \bibinfo{journal}{JHEP} \textbf{\bibinfo{volume}{09}}, \bibinfo{pages}{118}
  (\bibinfo{year}{2020}), \eprint{2004.00631}.

\bibitem[{\citenamefont{Mies et~al.}(2020)\citenamefont{Mies, Scherb, and
  Schwaller}}]{Mies:2020mzw}
\bibinfo{author}{\bibfnamefont{H.}~\bibnamefont{Mies}},
  \bibinfo{author}{\bibfnamefont{C.}~\bibnamefont{Scherb}}, \bibnamefont{and}
  \bibinfo{author}{\bibfnamefont{P.}~\bibnamefont{Schwaller}}
  (\bibinfo{year}{2020}), \eprint{2011.13990}.

\bibitem[{\citenamefont{Knapen et~al.}(2021)\citenamefont{Knapen, Shelton, and
  Xu}}]{Knapen:2021eip}
\bibinfo{author}{\bibfnamefont{S.}~\bibnamefont{Knapen}},
  \bibinfo{author}{\bibfnamefont{J.}~\bibnamefont{Shelton}}, \bibnamefont{and}
  \bibinfo{author}{\bibfnamefont{D.}~\bibnamefont{Xu}} (\bibinfo{year}{2021}),
  \eprint{2103.01238}.

\bibitem[{\citenamefont{Renner and Schwaller}(2018)}]{Renner:2018fhh}
\bibinfo{author}{\bibfnamefont{S.}~\bibnamefont{Renner}} \bibnamefont{and}
  \bibinfo{author}{\bibfnamefont{P.}~\bibnamefont{Schwaller}},
  \bibinfo{journal}{JHEP} \textbf{\bibinfo{volume}{08}}, \bibinfo{pages}{052}
  (\bibinfo{year}{2018}), \eprint{1803.08080}.

\bibitem[{\citenamefont{Alekhin et~al.}(2016)}]{Alekhin:2015byh}
\bibinfo{author}{\bibfnamefont{S.}~\bibnamefont{Alekhin}} \bibnamefont{et~al.},
  \bibinfo{journal}{Rept. Prog. Phys.} \textbf{\bibinfo{volume}{79}},
  \bibinfo{pages}{124201} (\bibinfo{year}{2016}), \eprint{1504.04855}.

\bibitem[{\citenamefont{Curtin et~al.}(2019)}]{Curtin:2018mvb}
\bibinfo{author}{\bibfnamefont{D.}~\bibnamefont{Curtin}} \bibnamefont{et~al.},
  \bibinfo{journal}{Rept. Prog. Phys.} \textbf{\bibinfo{volume}{82}},
  \bibinfo{pages}{116201} (\bibinfo{year}{2019}), \eprint{1806.07396}.

\bibitem[{\citenamefont{Sirunyan et~al.}(2019)}]{Sirunyan:2018njd}
\bibinfo{author}{\bibfnamefont{A.~M.} \bibnamefont{Sirunyan}}
  \bibnamefont{et~al.} (\bibinfo{collaboration}{CMS}), \bibinfo{journal}{JHEP}
  \textbf{\bibinfo{volume}{02}}, \bibinfo{pages}{179} (\bibinfo{year}{2019}),
  \eprint{1810.10069}.

\bibitem[{\citenamefont{Bai et~al.}(2010)\citenamefont{Bai, Fox, and
  Harnik}}]{Bai:2010hh}
\bibinfo{author}{\bibfnamefont{Y.}~\bibnamefont{Bai}},
  \bibinfo{author}{\bibfnamefont{P.~J.} \bibnamefont{Fox}}, \bibnamefont{and}
  \bibinfo{author}{\bibfnamefont{R.}~\bibnamefont{Harnik}},
  \bibinfo{journal}{JHEP} \textbf{\bibinfo{volume}{12}}, \bibinfo{pages}{048}
  (\bibinfo{year}{2010}), \eprint{1005.3797}.

\bibitem[{\citenamefont{Goodman et~al.}(2010)\citenamefont{Goodman, Ibe,
  Rajaraman, Shepherd, Tait, and Yu}}]{Goodman:2010ku}
\bibinfo{author}{\bibfnamefont{J.}~\bibnamefont{Goodman}},
  \bibinfo{author}{\bibfnamefont{M.}~\bibnamefont{Ibe}},
  \bibinfo{author}{\bibfnamefont{A.}~\bibnamefont{Rajaraman}},
  \bibinfo{author}{\bibfnamefont{W.}~\bibnamefont{Shepherd}},
  \bibinfo{author}{\bibfnamefont{T.~M.~P.} \bibnamefont{Tait}},
  \bibnamefont{and} \bibinfo{author}{\bibfnamefont{H.-B.} \bibnamefont{Yu}},
  \bibinfo{journal}{Phys. Rev. D} \textbf{\bibinfo{volume}{82}},
  \bibinfo{pages}{116010} (\bibinfo{year}{2010}), \eprint{1008.1783}.

\bibitem[{\citenamefont{Aad et~al.}(2021)}]{Aad:2021egl}
\bibinfo{author}{\bibfnamefont{G.}~\bibnamefont{Aad}} \bibnamefont{et~al.}
  (\bibinfo{collaboration}{ATLAS}) (\bibinfo{year}{2021}), \eprint{2102.10874}.

\bibitem[{\citenamefont{Sirunyan
  et~al.}(2017{\natexlab{a}})}]{Sirunyan:2017ewk}
\bibinfo{author}{\bibfnamefont{A.~M.} \bibnamefont{Sirunyan}}
  \bibnamefont{et~al.} (\bibinfo{collaboration}{CMS}), \bibinfo{journal}{JHEP}
  \textbf{\bibinfo{volume}{10}}, \bibinfo{pages}{073}
  (\bibinfo{year}{2017}{\natexlab{a}}), \eprint{1706.03794}.

\bibitem[{\citenamefont{Sirunyan
  et~al.}(2017{\natexlab{b}})}]{Sirunyan:2017hci}
\bibinfo{author}{\bibfnamefont{A.~M.} \bibnamefont{Sirunyan}}
  \bibnamefont{et~al.} (\bibinfo{collaboration}{CMS}), \bibinfo{journal}{JHEP}
  \textbf{\bibinfo{volume}{07}}, \bibinfo{pages}{014}
  (\bibinfo{year}{2017}{\natexlab{b}}), \eprint{1703.01651}.

\bibitem[{\citenamefont{Larkoski et~al.}(2020)\citenamefont{Larkoski, Moult,
  and Nachman}}]{Larkoski_2020}
\bibinfo{author}{\bibfnamefont{A.~J.} \bibnamefont{Larkoski}},
  \bibinfo{author}{\bibfnamefont{I.}~\bibnamefont{Moult}}, \bibnamefont{and}
  \bibinfo{author}{\bibfnamefont{B.}~\bibnamefont{Nachman}},
  \bibinfo{journal}{Phys. Rept.} \textbf{\bibinfo{volume}{841}},
  \bibinfo{pages}{1} (\bibinfo{year}{2020}), \eprint{1709.04464}.

\bibitem[{\citenamefont{Metodiev et~al.}(2017)\citenamefont{Metodiev, Nachman,
  and Thaler}}]{Metodiev_2017}
\bibinfo{author}{\bibfnamefont{E.~M.} \bibnamefont{Metodiev}},
  \bibinfo{author}{\bibfnamefont{B.}~\bibnamefont{Nachman}}, \bibnamefont{and}
  \bibinfo{author}{\bibfnamefont{J.}~\bibnamefont{Thaler}},
  \bibinfo{journal}{JHEP} \textbf{\bibinfo{volume}{10}}, \bibinfo{pages}{174}
  (\bibinfo{year}{2017}), \eprint{1708.02949}.

\bibitem[{\citenamefont{Metodiev and Thaler}(2018)}]{Metodiev_2018}
\bibinfo{author}{\bibfnamefont{E.~M.} \bibnamefont{Metodiev}} \bibnamefont{and}
  \bibinfo{author}{\bibfnamefont{J.}~\bibnamefont{Thaler}},
  \bibinfo{journal}{Phys. Rev. Lett.} \textbf{\bibinfo{volume}{120}},
  \bibinfo{pages}{241602} (\bibinfo{year}{2018}), \eprint{1802.00008}.

\bibitem[{\citenamefont{Guest et~al.}(2018)\citenamefont{Guest, Cranmer, and
  Whiteson}}]{doi:10.1146/annurev-nucl-101917-021019}
\bibinfo{author}{\bibfnamefont{D.}~\bibnamefont{Guest}},
  \bibinfo{author}{\bibfnamefont{K.}~\bibnamefont{Cranmer}}, \bibnamefont{and}
  \bibinfo{author}{\bibfnamefont{D.}~\bibnamefont{Whiteson}},
  \bibinfo{journal}{Ann. Rev. Nucl. Part. Sci.} \textbf{\bibinfo{volume}{68}},
  \bibinfo{pages}{161} (\bibinfo{year}{2018}), \eprint{1806.11484}.

\bibitem[{\citenamefont{Radovic et~al.}(2018)\citenamefont{Radovic, Williams,
  Rousseau, Kagan, Bonacorsi, Himmel, Aurisano, Terao, and
  Wongjirad}}]{Radovic}
\bibinfo{author}{\bibfnamefont{A.}~\bibnamefont{Radovic}},
  \bibinfo{author}{\bibfnamefont{M.}~\bibnamefont{Williams}},
  \bibinfo{author}{\bibfnamefont{D.}~\bibnamefont{Rousseau}},
  \bibinfo{author}{\bibfnamefont{M.}~\bibnamefont{Kagan}},
  \bibinfo{author}{\bibfnamefont{D.}~\bibnamefont{Bonacorsi}},
  \bibinfo{author}{\bibfnamefont{A.}~\bibnamefont{Himmel}},
  \bibinfo{author}{\bibfnamefont{A.}~\bibnamefont{Aurisano}},
  \bibinfo{author}{\bibfnamefont{K.}~\bibnamefont{Terao}}, \bibnamefont{and}
  \bibinfo{author}{\bibfnamefont{T.}~\bibnamefont{Wongjirad}},
  \bibinfo{journal}{Nature} \textbf{\bibinfo{volume}{560}}, \bibinfo{pages}{41}
  (\bibinfo{year}{2018}).

\bibitem[{\citenamefont{Alimena
  et~al.}(2020{\natexlab{b}})\citenamefont{Alimena, Iiyama, and
  Kieseler}}]{Alimena_2020}
\bibinfo{author}{\bibfnamefont{J.}~\bibnamefont{Alimena}},
  \bibinfo{author}{\bibfnamefont{Y.}~\bibnamefont{Iiyama}}, \bibnamefont{and}
  \bibinfo{author}{\bibfnamefont{J.}~\bibnamefont{Kieseler}},
  \bibinfo{journal}{JINST} \textbf{\bibinfo{volume}{15}}, \bibinfo{pages}{12}
  (\bibinfo{year}{2020}{\natexlab{b}}), \eprint{2004.10744}.

\bibitem[{\citenamefont{Gershtein et~al.}(2020)\citenamefont{Gershtein, Knapen,
  and Redigolo}}]{Gershtein:2020mwi}
\bibinfo{author}{\bibfnamefont{Y.}~\bibnamefont{Gershtein}},
  \bibinfo{author}{\bibfnamefont{S.}~\bibnamefont{Knapen}}, \bibnamefont{and}
  \bibinfo{author}{\bibfnamefont{D.}~\bibnamefont{Redigolo}}
  (\bibinfo{year}{2020}), \eprint{2012.07864}.

\bibitem[{\citenamefont{Mukherjee}(2020)}]{Mukherjee:2019anz}
\bibinfo{author}{\bibfnamefont{S.}~\bibnamefont{Mukherjee}}
  (\bibinfo{collaboration}{CMS}), \bibinfo{journal}{PoS}
  \textbf{\bibinfo{volume}{EPS-HEP2019}}, \bibinfo{pages}{139}
  (\bibinfo{year}{2020}).

\bibitem[{\citenamefont{Anderson}(2016)}]{Anderson:2016ron}
\bibinfo{author}{\bibfnamefont{D.}~\bibnamefont{Anderson}}
  (\bibinfo{collaboration}{CMS}), \bibinfo{journal}{PoS}
  \textbf{\bibinfo{volume}{ICHEP2016}}, \bibinfo{pages}{190}
  (\bibinfo{year}{2016}).

\bibitem[{ATL(2017)}]{ATL-DAQ-PUB-2017-003}
\bibinfo{type}{Tech. Rep.} \bibinfo{number}{ATL-DAQ-PUB-2017-003},
  \bibinfo{institution}{CERN}, \bibinfo{address}{Geneva}
  (\bibinfo{year}{2017}), \urlprefix\url{http://cds.cern.ch/record/2295739}.

\bibitem[{\citenamefont{Faucett et~al.}(2020)\citenamefont{Faucett, Thaler, and
  Whiteson}}]{Faucett:2020vbu}
\bibinfo{author}{\bibfnamefont{T.}~\bibnamefont{Faucett}},
  \bibinfo{author}{\bibfnamefont{J.}~\bibnamefont{Thaler}}, \bibnamefont{and}
  \bibinfo{author}{\bibfnamefont{D.}~\bibnamefont{Whiteson}}
  (\bibinfo{year}{2020}), \eprint{2010.11998}.

\bibitem[{\citenamefont{Owen}(2018)}]{Owen:2302730}
\bibinfo{author}{\bibfnamefont{R.~E.} \bibnamefont{Owen}}
  (\bibinfo{collaboration}{ATLAS Collaboration}) (\bibinfo{year}{2018}),
  \urlprefix\url{https://cds.cern.ch/record/2302730}.

\bibitem[{\citenamefont{Khachatryan et~al.}(2017)\citenamefont{Khachatryan,
  Sirunyan, Tumasyan, Adam, Asilar, Bergauer, Brandstetter, Brondolin,
  Dragicevic, Erö et~al.}}]{Khachatryan_2017}
\bibinfo{author}{\bibfnamefont{V.}~\bibnamefont{Khachatryan}},
  \bibinfo{author}{\bibfnamefont{A.}~\bibnamefont{Sirunyan}},
  \bibinfo{author}{\bibfnamefont{A.}~\bibnamefont{Tumasyan}},
  \bibinfo{author}{\bibfnamefont{W.}~\bibnamefont{Adam}},
  \bibinfo{author}{\bibfnamefont{E.}~\bibnamefont{Asilar}},
  \bibinfo{author}{\bibfnamefont{T.}~\bibnamefont{Bergauer}},
  \bibinfo{author}{\bibfnamefont{J.}~\bibnamefont{Brandstetter}},
  \bibinfo{author}{\bibfnamefont{E.}~\bibnamefont{Brondolin}},
  \bibinfo{author}{\bibfnamefont{M.}~\bibnamefont{Dragicevic}},
  \bibinfo{author}{\bibfnamefont{J.}~\bibnamefont{Erö}}, \bibnamefont{et~al.},
  \bibinfo{journal}{Journal of Instrumentation} \textbf{\bibinfo{volume}{12}},
  \bibinfo{pages}{P01020–P01020} (\bibinfo{year}{2017}), ISSN
  \bibinfo{issn}{1748-0221},
  \urlprefix\url{http://dx.doi.org/10.1088/1748-0221/12/01/P01020}.

\bibitem[{\citenamefont{Zyla et~al.}(2020)}]{Zyla:2020zbs}
\bibinfo{author}{\bibfnamefont{P.~A.} \bibnamefont{Zyla}} \bibnamefont{et~al.}
  (\bibinfo{collaboration}{Particle Data Group}), \bibinfo{journal}{PTEP}
  \textbf{\bibinfo{volume}{2020}}, \bibinfo{pages}{083C01}
  (\bibinfo{year}{2020}).

\bibitem[{\citenamefont{Witten}(1979)}]{Witten:1979kh}
\bibinfo{author}{\bibfnamefont{E.}~\bibnamefont{Witten}},
  \bibinfo{journal}{Nucl. Phys. B} \textbf{\bibinfo{volume}{160}},
  \bibinfo{pages}{57} (\bibinfo{year}{1979}).

\bibitem[{Phy(2017)}]{PhysRevD.96.052004}
\bibinfo{journal}{Phys. Rev. D} \textbf{\bibinfo{volume}{96}},
  \bibinfo{pages}{052004} (\bibinfo{year}{2017}),
  \urlprefix\url{https://link.aps.org/doi/10.1103/PhysRevD.96.052004}.

\bibitem[{\citenamefont{Aaboud et~al.}(2019{\natexlab{a}})}]{2019316}
\bibinfo{author}{\bibfnamefont{M.}~\bibnamefont{Aaboud}} \bibnamefont{et~al.}
  (\bibinfo{collaboration}{ATLAS}), \bibinfo{journal}{Phys. Lett. B}
  \textbf{\bibinfo{volume}{788}}, \bibinfo{pages}{316}
  (\bibinfo{year}{2019}{\natexlab{a}}), \eprint{1801.08769}.

\bibitem[{\citenamefont{Sirunyan et~al.}(2017{\natexlab{c}})}]{Sirunyan_2019}
\bibinfo{author}{\bibfnamefont{A.~M.} \bibnamefont{Sirunyan}}
  \bibnamefont{et~al.} (\bibinfo{collaboration}{CMS}), \bibinfo{journal}{Phys.
  Rev. Lett.} \textbf{\bibinfo{volume}{119}}, \bibinfo{pages}{111802}
  (\bibinfo{year}{2017}{\natexlab{c}}), \eprint{1705.10532}.

\bibitem[{\citenamefont{Sirunyan et~al.}(2017{\natexlab{d}})}]{2017520}
\bibinfo{author}{\bibfnamefont{A.~M.} \bibnamefont{Sirunyan}}
  \bibnamefont{et~al.} (\bibinfo{collaboration}{CMS}), \bibinfo{journal}{Phys.
  Lett. B} \textbf{\bibinfo{volume}{769}}, \bibinfo{pages}{520}
  (\bibinfo{year}{2017}{\natexlab{d}}), \bibinfo{note}{[Erratum: Phys.Lett.B
  772, 882--883 (2017)]}, \eprint{1611.03568}.

\bibitem[{Phy(2018)}]{PhysRevLett.121.081801}
\bibinfo{journal}{Phys. Rev. Lett.} \textbf{\bibinfo{volume}{121}},
  \bibinfo{pages}{081801} (\bibinfo{year}{2018}),
  \urlprefix\url{https://link.aps.org/doi/10.1103/PhysRevLett.121.081801}.

\bibitem[{\citenamefont{Alloul et~al.}(2014)\citenamefont{Alloul, Christensen,
  Degrande, Duhr, and Fuks}}]{Alloul_2014}
\bibinfo{author}{\bibfnamefont{A.}~\bibnamefont{Alloul}},
  \bibinfo{author}{\bibfnamefont{N.~D.} \bibnamefont{Christensen}},
  \bibinfo{author}{\bibfnamefont{C.}~\bibnamefont{Degrande}},
  \bibinfo{author}{\bibfnamefont{C.}~\bibnamefont{Duhr}}, \bibnamefont{and}
  \bibinfo{author}{\bibfnamefont{B.}~\bibnamefont{Fuks}},
  \bibinfo{journal}{Comput. Phys. Commun.} \textbf{\bibinfo{volume}{185}},
  \bibinfo{pages}{2250} (\bibinfo{year}{2014}), \eprint{1310.1921}.

\bibitem[{\citenamefont{Degrande et~al.}(2012)\citenamefont{Degrande, Duhr,
  Fuks, Grellscheid, Mattelaer, and Reiter}}]{Degrande_2012}
\bibinfo{author}{\bibfnamefont{C.}~\bibnamefont{Degrande}},
  \bibinfo{author}{\bibfnamefont{C.}~\bibnamefont{Duhr}},
  \bibinfo{author}{\bibfnamefont{B.}~\bibnamefont{Fuks}},
  \bibinfo{author}{\bibfnamefont{D.}~\bibnamefont{Grellscheid}},
  \bibinfo{author}{\bibfnamefont{O.}~\bibnamefont{Mattelaer}},
  \bibnamefont{and} \bibinfo{author}{\bibfnamefont{T.}~\bibnamefont{Reiter}},
  \bibinfo{journal}{Comput. Phys. Commun.} \textbf{\bibinfo{volume}{183}},
  \bibinfo{pages}{1201} (\bibinfo{year}{2012}), \eprint{1108.2040}.

\bibitem[{\citenamefont{Alwall et~al.}(2014)\citenamefont{Alwall, Frederix,
  Frixione, Hirschi, Maltoni, Mattelaer, Shao, Stelzer, Torrielli, and
  Zaro}}]{Alwall:2014hca}
\bibinfo{author}{\bibfnamefont{J.}~\bibnamefont{Alwall}},
  \bibinfo{author}{\bibfnamefont{R.}~\bibnamefont{Frederix}},
  \bibinfo{author}{\bibfnamefont{S.}~\bibnamefont{Frixione}},
  \bibinfo{author}{\bibfnamefont{V.}~\bibnamefont{Hirschi}},
  \bibinfo{author}{\bibfnamefont{F.}~\bibnamefont{Maltoni}},
  \bibinfo{author}{\bibfnamefont{O.}~\bibnamefont{Mattelaer}},
  \bibinfo{author}{\bibfnamefont{H.~S.} \bibnamefont{Shao}},
  \bibinfo{author}{\bibfnamefont{T.}~\bibnamefont{Stelzer}},
  \bibinfo{author}{\bibfnamefont{P.}~\bibnamefont{Torrielli}},
  \bibnamefont{and} \bibinfo{author}{\bibfnamefont{M.}~\bibnamefont{Zaro}},
  \bibinfo{journal}{JHEP} \textbf{\bibinfo{volume}{07}}, \bibinfo{pages}{079}
  (\bibinfo{year}{2014}), \eprint{1405.0301}.

\bibitem[{\citenamefont{Carloni and
  Sjostrand}(2010{\natexlab{a}})}]{Carloni_2010}
\bibinfo{author}{\bibfnamefont{L.}~\bibnamefont{Carloni}} \bibnamefont{and}
  \bibinfo{author}{\bibfnamefont{T.}~\bibnamefont{Sjostrand}},
  \bibinfo{journal}{JHEP} \textbf{\bibinfo{volume}{09}}, \bibinfo{pages}{105}
  (\bibinfo{year}{2010}{\natexlab{a}}), \eprint{1006.2911}.

\bibitem[{\citenamefont{Carloni and
  Sjostrand}(2010{\natexlab{b}})}]{Carloni_2011}
\bibinfo{author}{\bibfnamefont{L.}~\bibnamefont{Carloni}} \bibnamefont{and}
  \bibinfo{author}{\bibfnamefont{T.}~\bibnamefont{Sjostrand}},
  \bibinfo{journal}{JHEP} \textbf{\bibinfo{volume}{09}}, \bibinfo{pages}{105}
  (\bibinfo{year}{2010}{\natexlab{b}}), \eprint{1006.2911}.

\bibitem[{\citenamefont{Sjöstrand et~al.}(2015)\citenamefont{Sjöstrand, Ask,
  Christiansen, Corke, Desai, Ilten, Mrenna, Prestel, Rasmussen, and
  Skands}}]{Sjostrand:2014zea}
\bibinfo{author}{\bibfnamefont{T.}~\bibnamefont{Sjöstrand}},
  \bibinfo{author}{\bibfnamefont{S.}~\bibnamefont{Ask}},
  \bibinfo{author}{\bibfnamefont{J.~R.} \bibnamefont{Christiansen}},
  \bibinfo{author}{\bibfnamefont{R.}~\bibnamefont{Corke}},
  \bibinfo{author}{\bibfnamefont{N.}~\bibnamefont{Desai}},
  \bibinfo{author}{\bibfnamefont{P.}~\bibnamefont{Ilten}},
  \bibinfo{author}{\bibfnamefont{S.}~\bibnamefont{Mrenna}},
  \bibinfo{author}{\bibfnamefont{S.}~\bibnamefont{Prestel}},
  \bibinfo{author}{\bibfnamefont{C.~O.} \bibnamefont{Rasmussen}},
  \bibnamefont{and} \bibinfo{author}{\bibfnamefont{P.~Z.}
  \bibnamefont{Skands}}, \bibinfo{journal}{Comput. Phys. Commun.}
  \textbf{\bibinfo{volume}{191}}, \bibinfo{pages}{159} (\bibinfo{year}{2015}),
  \eprint{1410.3012}.

\bibitem[{\citenamefont{Mangano et~al.}(2007)\citenamefont{Mangano, Moretti,
  Piccinini, and Treccani}}]{Mangano_2007}
\bibinfo{author}{\bibfnamefont{M.~L.} \bibnamefont{Mangano}},
  \bibinfo{author}{\bibfnamefont{M.}~\bibnamefont{Moretti}},
  \bibinfo{author}{\bibfnamefont{F.}~\bibnamefont{Piccinini}},
  \bibnamefont{and} \bibinfo{author}{\bibfnamefont{M.}~\bibnamefont{Treccani}},
  \bibinfo{journal}{JHEP} \textbf{\bibinfo{volume}{01}}, \bibinfo{pages}{013}
  (\bibinfo{year}{2007}), \eprint{hep-ph/0611129}.

\bibitem[{\citenamefont{Cacciari et~al.}(2008)\citenamefont{Cacciari, Salam,
  and Soyez}}]{Cacciari_2008}
\bibinfo{author}{\bibfnamefont{M.}~\bibnamefont{Cacciari}},
  \bibinfo{author}{\bibfnamefont{G.~P.} \bibnamefont{Salam}}, \bibnamefont{and}
  \bibinfo{author}{\bibfnamefont{G.}~\bibnamefont{Soyez}},
  \bibinfo{journal}{JHEP} \textbf{\bibinfo{volume}{04}}, \bibinfo{pages}{063}
  (\bibinfo{year}{2008}), \eprint{0802.1189}.

\bibitem[{\citenamefont{Cacciari et~al.}(2012)\citenamefont{Cacciari, Salam,
  and Soyez}}]{Cacciari:2011ma}
\bibinfo{author}{\bibfnamefont{M.}~\bibnamefont{Cacciari}},
  \bibinfo{author}{\bibfnamefont{G.~P.} \bibnamefont{Salam}}, \bibnamefont{and}
  \bibinfo{author}{\bibfnamefont{G.}~\bibnamefont{Soyez}},
  \bibinfo{journal}{Eur. Phys. J.} \textbf{\bibinfo{volume}{C72}},
  \bibinfo{pages}{1896} (\bibinfo{year}{2012}), \eprint{1111.6097}.

\bibitem[{ATL(2018)}]{ATL-DAQ-PUB-2018-002}
\bibinfo{type}{Tech. Rep.} \bibinfo{number}{ATL-DAQ-PUB-2018-002},
  \bibinfo{institution}{CERN}, \bibinfo{address}{Geneva}
  (\bibinfo{year}{2018}), \urlprefix\url{https://cds.cern.ch/record/2625986}.

\bibitem[{\citenamefont{Aad et~al.}(2020)\citenamefont{Aad, Abbott, Abbott,
  Abud, Abeling, Abhayasinghe, Abidi, AbouZeid, Abraham, and
  et~al.}}]{Aad_2020}
\bibinfo{author}{\bibfnamefont{G.}~\bibnamefont{Aad}},
  \bibinfo{author}{\bibfnamefont{B.}~\bibnamefont{Abbott}},
  \bibinfo{author}{\bibfnamefont{D.~C.} \bibnamefont{Abbott}},
  \bibinfo{author}{\bibfnamefont{A.~A.} \bibnamefont{Abud}},
  \bibinfo{author}{\bibfnamefont{K.}~\bibnamefont{Abeling}},
  \bibinfo{author}{\bibfnamefont{D.~K.} \bibnamefont{Abhayasinghe}},
  \bibinfo{author}{\bibfnamefont{S.~H.} \bibnamefont{Abidi}},
  \bibinfo{author}{\bibfnamefont{O.~S.} \bibnamefont{AbouZeid}},
  \bibinfo{author}{\bibfnamefont{N.~L.} \bibnamefont{Abraham}},
  \bibnamefont{and} \bibinfo{author}{\bibnamefont{et~al.}},
  \bibinfo{journal}{The European Physical Journal C}
  \textbf{\bibinfo{volume}{80}} (\bibinfo{year}{2020}), ISSN
  \bibinfo{issn}{1434-6052},
  \urlprefix\url{http://dx.doi.org/10.1140/epjc/s10052-019-7500-2}.

\bibitem[{\citenamefont{Aaboud et~al.}(2019{\natexlab{b}})}]{Aaboud_2019}
\bibinfo{author}{\bibfnamefont{M.}~\bibnamefont{Aaboud}} \bibnamefont{et~al.}
  (\bibinfo{collaboration}{ATLAS}), \bibinfo{journal}{Eur. Phys. J. C}
  \textbf{\bibinfo{volume}{79}}, \bibinfo{pages}{205}
  (\bibinfo{year}{2019}{\natexlab{b}}), \eprint{1810.05087}.

\bibitem[{Air(1999)}]{Airapetian:391176}
\emph{\bibinfo{title}{{ATLAS detector and physics performance: Technical Design
  Report, 1}}}, Technical design report. ATLAS (\bibinfo{publisher}{CERN},
  \bibinfo{address}{Geneva}, \bibinfo{year}{1999}),
  \urlprefix\url{https://cds.cern.ch/record/391176}.

\bibitem[{\citenamefont{Capeans et~al.}(2010)\citenamefont{Capeans, Darbo,
  Einsweiller, Elsing, Flick, Garcia-Sciveres, Gemme, Pernegger, Rohne, and
  Vuillermet}}]{Capeans:1291633}
\bibinfo{author}{\bibfnamefont{M.}~\bibnamefont{Capeans}},
  \bibinfo{author}{\bibfnamefont{G.}~\bibnamefont{Darbo}},
  \bibinfo{author}{\bibfnamefont{K.}~\bibnamefont{Einsweiller}},
  \bibinfo{author}{\bibfnamefont{M.}~\bibnamefont{Elsing}},
  \bibinfo{author}{\bibfnamefont{T.}~\bibnamefont{Flick}},
  \bibinfo{author}{\bibfnamefont{M.}~\bibnamefont{Garcia-Sciveres}},
  \bibinfo{author}{\bibfnamefont{C.}~\bibnamefont{Gemme}},
  \bibinfo{author}{\bibfnamefont{H.}~\bibnamefont{Pernegger}},
  \bibinfo{author}{\bibfnamefont{O.}~\bibnamefont{Rohne}}, \bibnamefont{and}
  \bibinfo{author}{\bibfnamefont{R.}~\bibnamefont{Vuillermet}}
  (\bibinfo{collaboration}{ATLAS Collaboration}), \bibinfo{type}{Tech. Rep.}
  \bibinfo{number}{CERN-LHCC-2010-013. ATLAS-TDR-19} (\bibinfo{year}{2010}),
  \urlprefix\url{https://cds.cern.ch/record/1291633}.

\bibitem[{\citenamefont{Hoecker et~al.}(2007)\citenamefont{Hoecker, Speckmayer,
  Stelzer, Therhaag, von Toerne, and Voss}}]{Hocker:2007ht}
\bibinfo{author}{\bibfnamefont{A.}~\bibnamefont{Hoecker}},
  \bibinfo{author}{\bibfnamefont{P.}~\bibnamefont{Speckmayer}},
  \bibinfo{author}{\bibfnamefont{J.}~\bibnamefont{Stelzer}},
  \bibinfo{author}{\bibfnamefont{J.}~\bibnamefont{Therhaag}},
  \bibinfo{author}{\bibfnamefont{E.}~\bibnamefont{von Toerne}},
  \bibnamefont{and} \bibinfo{author}{\bibfnamefont{H.}~\bibnamefont{Voss}},
  \bibinfo{journal}{PoS} \textbf{\bibinfo{volume}{ACAT}}, \bibinfo{pages}{040}
  (\bibinfo{year}{2007}), \eprint{physics/0703039}.

\bibitem[{\citenamefont{Aad et~al.}(2013)}]{2013LLPTrigger}
\bibinfo{author}{\bibfnamefont{G.}~\bibnamefont{Aad}} \bibnamefont{et~al.}
  (\bibinfo{collaboration}{ATLAS}), \bibinfo{journal}{JINST}
  \textbf{\bibinfo{volume}{8}}, \bibinfo{pages}{P07015} (\bibinfo{year}{2013}),
  \eprint{1305.2284}.

\bibitem[{\citenamefont{Cesarotti and Thaler}(2020)}]{Cesarotti:2020hwb}
\bibinfo{author}{\bibfnamefont{C.}~\bibnamefont{Cesarotti}} \bibnamefont{and}
  \bibinfo{author}{\bibfnamefont{J.}~\bibnamefont{Thaler}},
  \bibinfo{journal}{JHEP} \textbf{\bibinfo{volume}{08}}, \bibinfo{pages}{084}
  (\bibinfo{year}{2020}), \eprint{2004.06125}.

\end{thebibliography}

\end{document}